\documentclass{aa}

\usepackage{graphicx}
\usepackage{txfonts}
\usepackage{natbib}
\bibpunct{(}{)}{;}{a}{}{,} 

\usepackage{hyperref}
\hypersetup{
    colorlinks=true,
    citecolor=blue,
    linkcolor=blue,
    filecolor=magenta,      
    urlcolor=blue,
}

\begin{document}

\title{Science Using Single-Pulse Exploration with Combined Telescopes}
\subtitle{I. The mode switching, flaring, and single-pulse morphology of PSR~B1822$-$09}
\titlerunning{Science Using Single-Pulse Exploration with Combined Telescopes I}

\author{
F.~Jankowski\inst{1}\thanks{Corresponding author; \texttt{fabian.jankowski@cnrs-orleans.fr}}
\and
J.-M.~Grie{\ss}meier\inst{1,2}
\and
M.~Surnis\inst{3}
\and
G.~Theureau\inst{1,2,4}
\and
J.~P\'etri\inst{5}
}

\institute{
LPC2E, OSUC, Univ Orleans, CNRS, CNES, Observatoire de Paris, F-45071 Orleans, France
\and
Observatoire Radioastronomique de Nan\c{c}ay, Observatoire de Paris, Universit\'e PSL, Université d'Orl\'eans, CNRS, 18330 Nan\c{c}ay, France
\and
Department of Physics, IISER Bhopal, Bhauri Bypass Road, Bhopal, 462066, India
\and
LUTH, Observatoire de Paris, Universit\'e PSL, Universit\'e Paris Cit\'e, CNRS, 92195 Meudon, France
\and
Universit\'e de Strasbourg, CNRS, Observatoire Astronomique de Strasbourg, UMR 7550, F-67000 Strasbourg, France
}

\date{Received XXX; accepted XXX}

 
\abstract
{
Radio pulsars exhibit a plethora of complex phenomena at the single-pulse level. However, the intricacies of their radio emission remain poorly understood.
}
{
We aim to elucidate the pulsar radio emission by studying several single-pulse phenomena, how they relate, and how they evolve with observing frequency. We intend to inspire models for the pulsar radio emission and fast radio bursts.
}
{
We set up an observing programme called the SUSPECT project running at the Nan\c{c}ay Radio Observatory telescopes in France (10--85~MHz, 110--240~MHz, and 1.1--3.5~GHz) and the upgraded Giant Metrewave Radio Telescope (uGMRT) in India. This first paper focuses on high sensitivity data of PSR~B1822$-$09 obtained with the uGMRT between 550 and 750~MHz. The pulsar has precursor (PC), main pulse (MP), and interpulse (IP) emission and exhibits mode switching. We present its single-pulse stacks, investigate its mode switching using a hidden Markov switching model, and analyse its single-pulse morphology.
}
{
PSR~B1822$-$09's pulse profile decomposes into seven components. We show that its mode switching is well described using a hidden Markov switching model operating on single-pulse profile features. The pulsar exhibits at least three stable emission modes, one of which is a newly discovered bright flaring Bf-mode. We confirm that the PC and MP switch synchronously to each other and both asynchronously to the IP, indicating information transfer between the polar caps. Additionally, we performed a fluctuation spectral analysis and discovered three fluctuation features in its quiescent Q-mode emission, one of which is well known. We conclude that the latter feature is due to longitude-stationary amplitude modulation. Finally, we visually classified the single pulses into four categories. We found extensive microstructure in the PC with a typical duration of 0.2--0.4~ms and a quasi-periodicity of 0.8~ms. There is clear evidence of mode mixing. We discovered low-intensity square-like pulses and extremely bright pulses in the MP, which suggest bursting.
}
{
PSR~B1822$-$09's PC resembles magnetar radio emission, while its MP and IP are canonical radio pulsar-like. Hence, the pulsar combines both attributes, which is rare. This work introduces several new data analysis techniques to pulsar astrophysics.
}


\maketitle
%

\section{Introduction}
\label{sec:introduction}

The vast majority of radio pulsars exhibit pulse profiles that are stable in time when thousands of their individual pulses are averaged. However, a small number of pulsars are known to switch between two or more stable emission modes with distinct pulse profiles. This phenomenon is commonly known as mode switching \citep{1970BackerModeChanging}. The switching between the emission modes can happen quasi-randomly and is in some cases correlated with changes in the spin-down rate $\dot{\nu}$ of the neutron star \citep{2010Lyne}. A closely related phenomenon is the apparent cessation of radio emission from a pulsar, known as nulling \citep{1970BackerNulling, 1992Biggs, 2007Wang}. A more extreme case is exhibited by the so-called long-term intermittent pulsars that routinely turn off for weeks to years \citep{2006Kramer, 2017Lyne}. Other pulsars show systematic and recurring shifts of their single-pulse profiles in pulse phase from rotation to rotation, known as drifting sub-pulses \citep{2006Weltevrede, 2023Song}. While drifting sub-pulses seem to be relatively common in the pulsar population \citep{1973Backer, 2006Weltevrede, 2023Song}, nulling, mode switching, and long-term intermittency seem to be much rarer phenomena \citep{1981Fowler}. The list above is ordered by decreasing relative occurrence, which is influenced by observational bias from insufficient monitoring of large pulsar samples. In any case, the fact that only a small number of pulsars (a few percent of the current population) are known to exhibit mode switching and other related phenomena makes them attractive exceptional cases. We argue that they provide essential clues to understanding the physical properties and effects in the highly magnetised plasma that fills pulsar magnetospheres and the radio emission mechanism.


\begin{figure}
  \centering
  \includegraphics[width=\columnwidth]{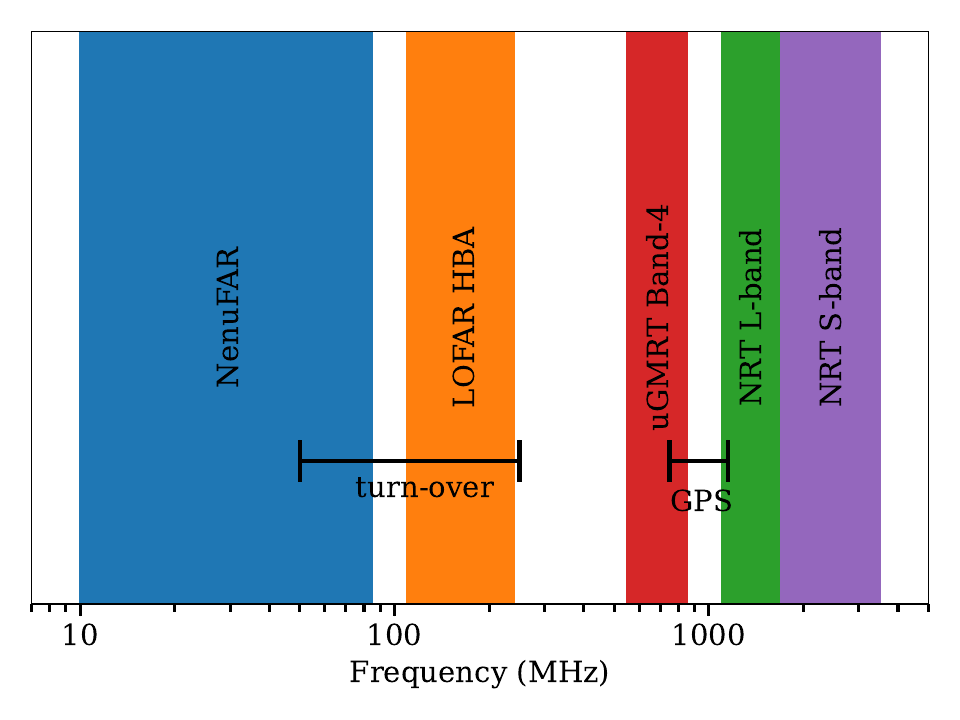}
  \caption{Schematic overview illustrating the frequency coverage of our data from the Nan\c{c}ay Radio Observatory telescopes (NenuFAR, LOFAR FR606, and NRT) together with the uGMRT Band-4 data. We also show the typical frequency ranges of low-frequency turn-overs in pulsar radio spectra and the peak frequencies of gigahertz-peaked spectrum (GPS) pulsars for reference.}
 \label{fig:freqcoverage}
\end{figure}

We devised an observing programme called the `Science Using Single-Pulse Exploration with Combined Telescopes' (SUSPECT) project\footnote{\url{https://suspectproject.com}} to better understand several pulsar single-pulse phenomena. The acronym also encapsulates our focus on several of the `usual suspect' pulsars, that is, widely known, relatively bright radio pulsars to achieve the required high signal-to-noise ratio (S/N) per pulse. The words `combined telescopes' indicate that we acquire most of our data using multi-element radio interferometers. Our primary objectives are to (1) measure and analyse the frequency dependence of pulsar single-pulse properties, such as mode switching, nulling, and sub-pulse drifting. (2) Secondly, we focus on exotic single-pulse behaviour, such as flaring, bursting, or `swooshing' \citep{2006Rankin, 2016Wahl, 2021Basu}. (3) Another motivation is to relate our pulsar single-pulse measurements with fast radio burst (FRB) phenomenology as observed, for instance, in repeating FRBs. We aim to inspire models for the FRB emission based on the plasma physics processes happening in pulsar magnetospheres, which themselves are not fully understood. Our more traditional objectives are (4) to study the evolution of the pulsars' integrated pulse profiles with frequency and to distinguish between propagation and intrinsic effects, such as the radius-to-frequency mapping model \citep{1978Cordes, 2016Pilia, 2021Posselt}. In this paradigm, the radio emission observed at different frequencies is believed to originate from different altitudes in the pulsar's magnetosphere. We also aim to (5) measure the pulsars' radio spectra with high accuracy, as the simultaneous wide frequency coverage of our data naturally allows this. A solid connection to the absolute flux density scale of the primary (gain, phase, and flux) calibrator from synthesis imaging is incredibly beneficial for this endeavour.

To study the objectives above, it is crucial to obtain pulsar measurements over a wide frequency range, ideally truly simultaneously. This is motivated by a desire to eliminate the effect of temporal changes in the pulsar intrinsic emission process and any propagation effects in the interstellar medium. Pulsar data collected simultaneously ensure that the telescopes' observing frequency is the only variable, the pulsar was in precisely the same emission state, and the interstellar medium remained almost unchanged.\footnote{That is, to a good first order. However, the Earth's ionosphere induces location-dependent effects that must ideally be compensated.} In practice, we achieved this by observing our pulsar sample simultaneously with several telescopes at the Nan\c{c}ay Radio Observatory in France, each covering a distinct but highly complementary frequency range. For additional frequency coverage, we observed with the upgraded Giant Metrewave Radio Telescope (uGMRT) in India. In particular, we used the New Extension in Nan\c{c}ay Upgrading LOFAR (NenuFAR) telescope (10--85~MHz), the HBA antennas of the French LOw-Frequency ARray (LOFAR) station FR606 (110--240~MHz), the uGMRT in Band-4 (550--750~MHz), and the Nan\c{c}ay Radio Telescope (NRT) at 1.1--1.8 and 1.7--3.5~GHz. Figure~\ref{fig:freqcoverage} visualises the frequency coverage schematically. We might extend the project to other facilities in the future.


\object{PSR~B1822$-$09} (\object{PSR~J1825$-$0935}) was discovered in the Jodrell Bank A pulsar survey at 408~MHz \citep{1972Davies} and is an interesting pulsar that exhibits a wealth of peculiar phenomena. It is one of the rare pulsars with interpulse emission separated by about half a turn from the main pulse that is most likely coming from the opposite pole of the star \citep{2017Hermsen}. Moreover, it shows mode switching, sub-pulse drifting \citep{1981Fowler}, spin-up glitches with strange recoveries \citep{2007Shabanova, 2010Yuan}, and correlated profile and spin-down rate changes \citep{2010Lyne}. It is a relatively young 769-ms period radio pulsar with an unusually high period derivative of $5.24 \times 10^{-14} \: \text{s} \text{s}^{-1}$ \citep{2019Jankowski}. It has a dispersion measure (DM) of 19.3833(9)~$\text{pc} \: \text{cm}^{-3}$ \citep{2015Stovall}, which places it at a DM-inferred distance of 262~pc, according to the \texttt{YMW16} Galactic free-electron model \citep{2017Yao}. Additionally, based on an HI emission or absorption distance limit, its Lutz-Kelker bias corrected distance is $0.3^{+0.7}_{-0.2}~\text{kpc}$ \citep{2012Verbiest}. It has a characteristic age of 233~kyr, a spin-down luminosity of $4.5 \times 10^{33}~\text{erg} \: \text{s}^{-1}$, and a surface dipole magnetic field strength of $6.42 \times 10^{12}~\text{G}$. The pulsar is reasonably bright in the radio band, with a mode-averaged flux density of 24~mJy at 843~MHz \citep{2019Jankowski} and 10~mJy at 1.4~GHz \citep{2018Johnston}. Its radio spectrum is highly complex \citep{1994Gil}, with published detections down to 42~MHz \citep{2012Suleymanova} and below 40~MHz with NenuFAR (private communication). An associated X-ray counterpart was identified \citep{2015Prinz}, and its sinusoidal pulsed emission was unambiguously detected \citep{2017Hermsen}. Gamma-ray counterparts were suggested early on \citep{1979Pinkau, 1980Mandrou}, and \citet{2017Hermsen} presented a marginal detection of the pulsar's emission in GeV $\gamma$-rays based on \textit{Fermi} satellite data. However, the pulsar is absent in the third \textit{Fermi} Large Area Telescope Pulsar Catalogue \citep{2023Smith}.

PSR~B1822$-$09 is a well-known mode changing pulsar that exhibits classical mode switching with two distinct emission modes. Its radio pulse profile consists of a main pulse (MP), a precursor (PC), and an interpulse (IP) profile component \citep{1981Fowler, 1981Morris}. It has a relatively stable and long-lasting quiescent mode (Q-mode) of typically 4.5-min duration that is present about 64\% of the time. It is interrupted by a significantly brighter burst mode (B-mode) that is short-lived and typically lasts only 2.5~min \citep{1981Fowler, 2017Hermsen}. The B-mode is characterised by the appearance of the PC component about 15~deg before the MP, which is absent or extremely weak in the Q-mode. The MP consists of two sub-components with a characteristic shoulder \citep{1981Fowler, 1994Gil}. The pulsar exhibits mode-dependent sub-pulse drifting and modulation. During the Q-mode, the MP and IP show a long-period modulation with a periodicity $P_3 = 40 - 47 \: P_1$, where $P_1$ is the pulsar's rotation period. This was initially thought to be caused by sub-pulse drifting \citep{1981Fowler}, but it seems to be due to longitude-stationary intensity modulation of unknown physical origin \citep{1994Gil, 2019Yan} or a mixture of amplitude and phase modulation \citep{2010Backus, 2012Latham}. In the B-mode, the long-period modulation ceases, and only short sequences of pulses with a much faster drift rate of $P_3 \approx 11 \: P_1$ have been reported \citep{1981Fowler, 1994Gil}. In data obtained at 325~MHz, \citet{2012Latham} reported a modulation of $P_3 = 70 \: P_1$. Interestingly, the PC component appears to have an almost flat spectral index, which is significantly different from that of the MP \citep{1981Fowler, 1994Gil}. The PC is almost 100\% linearly polarised and shows a flat polarisation position angle reminiscent of the Crab pulsar's precursor, which has a very steep radio spectrum \citep{1980Manchester, 2018Johnston}. Despite the detection of pulsed X-ray emission from the pulsar, no evidence was found for simultaneous mode switching between its X-ray and radio emission \citep{2017Hermsen}, which is in contrast to what has been discovered for other mode switching radio and X-ray pulsars, although the sample size is tiny (three; \citealt{2018Hermsen}). Moreover, PSR~B1822$-$09 is one of the few pulsars for which \citet{2010Lyne} found a significant correlation between its profile shape and spin-down rate. The ratio of its PC and MP amplitudes correlates positively with its spin-down rate on $\sim$10~yr timescales \citep{2010Lyne}.

Inferring a pulsar's true geometry is challenging. However, the most prevalent belief is that PSR~B1822$-$09 is an almost orthogonal rotator and that the PC and MP on the one hand and the IP on the other come from opposing magnetic poles of the star \citep{1986Hankins, 1994Gil, 2010Backus, 2017Hermsen}. Curiously, the IP participates in the mode changing, switching intensity asynchronously with the PC and synchronously with the MP \citep{1994Gil}. This phase-locked relationship between the emission from both poles of the star strongly suggests an information transfer between the radio emission regions at opposite poles, which is unexpected and challenging to explain physically \citep{1981Fowler, 1982Fowler, 1994Gil}. \citet{2005DyksA} suggested a somewhat unorthodox model in which a single radiation source creates both the PC and IP. They proposed that the IP originates because of a 180-deg flip in the emission direction of the source inwards, that is, towards the star.

PSR~B1822$-$09's rotation is regularly interrupted by spin-up glitches, with seven events currently listed in the Jodrell Bank glitch catalogue \citep{2011Espinoza} and 14 in the Australia Telescope National Facility pulsar catalogue and glitch table \citep{2005Manchester}. Several of them are so-called slow glitches with unusual signatures that consist of sharp increases in spin-down rate $\dot{\nu}$ that relax quasi-exponentially to their pre-glitch values over hundreds of days \citep{2007Shabanova, 2010Yuan, 2010Lyne}. The two strongest $\dot{\nu}$ events appear to be positively correlated with the pulse shape, namely the PC to MP amplitude ratio \citep{2010Lyne}. That is, the amplitude ratio PC / MP increased during episodes of increased $\dot{\nu}$. Interestingly, there is some evidence that its integrated pulse profile in both emission modes changed subtly after two glitches of normal signature \citep{2022Liu}.

In Sect.~\ref{sec:observations}, we describe our observations and the data reduction performed. In Sect.~\ref{sec:results}, we present the pulsar's integrated pulse profile, the single-pulse stacks, our model for its mode switching process, mode-separated profiles, our fluctuation spectral analysis, and a detailed analysis of its single-pulse morphology. In Sect.~\ref{sec:discussion}, we discuss our results and compare them with the literature. Finally, we present our conclusions in Sect.~\ref{sec:conclusions}.

\section{Observations and data processing}
\label{sec:observations}

\begin{table*}
\caption{Details of the uGMRT data presented in this work.}
\label{tab:observations}
\begin{tabular}{lccccccccc}
\hline\hline
Date            & Start UT      & $t_\text{obs}$\tablefootmark{a}    & $\nu_\text{c}$\tablefootmark{b}    & $b$\tablefootmark{c}   & $N_\text{chan}$\tablefootmark{d}   & $t_\text{samp}$\tablefootmark{e}   & $N_\text{ant,im}$\tablefootmark{f} & $N_\text{ant,pa}$\tablefootmark{g}    & $N_\text{pulse}$\tablefootmark{h}\\
(yyyy-mm-dd)    & (hh:mm)       & (min)             & (MHz)             & (MHz) &   & ($\mu \text{s}$)   &   &   &\\
\hline
2023-04-24      & 22:22         & 56 + 56           & 650               & 200   & 2048  & 81.92 & 29    & 16    & 8743\\
2023-12-05      & 11:40         & 45 + 37           & 650               & 200   & 2048  & 81.92 & 28    & 16    & 6362\\
\hline
Total           & --            & 194               & --                & --    & --    & --    & --    & --    & 15\,105\\
\hline
\end{tabular}
\tablefoot{
    \tablefoottext{a}{Observing time.}
    \tablefoottext{b}{Centre frequency.}
    \tablefoottext{c}{Digitised bandwidth.}
    \tablefoottext{d}{Number of frequency channels.}
    \tablefoottext{e}{Sampling time.}
    \tablefoottext{f}{Number of antennas in the imaging data stream.}
    \tablefoottext{g}{Number of antennas in the phased array data stream.}
    \tablefoottext{h}{Number of pulsar single pulses recorded.}
}
\end{table*}

Although the SUSPECT project is a multi-telescope, multi-frequency observing programme, we focus on high sensitivity single-band GMRT data in this first publication for brevity reasons. As part of the uGMRT project `Understanding the wide-band single-pulse properties of bright radio pulsars with the upgraded GMRT' (44\_056 and 45\_029, PI: Jankowski), we obtained beamformed high time-resolution data of PSR~B1822$-$09 at Band-4 frequencies. Starting from UT~2023-04-24 22:22, we recorded $2 \times 56~\text{min}$ of total intensity (Stokes I) data in two pointings separated by a phase calibrator scan of 6~min. On UT~2023-12-05 11:40, we observed the pulsar for 45 and 37~min again separated by a 6~min calibrator scan. The data were obtained between 550--750~MHz with 200~MHz of digitised bandwidth. We used the GMRT Wideband Backend (GWB) to record 2048 frequency channels at a sampling time of $81.92~\mu\text{s}$. Both 8-bit phased-array (PA) and 16-bit data coherently dedispersed (CD) at the pulsar's catalogued DM of 19.3833~$\text{pc} \: \text{cm}^{-3}$ \citep{2015Stovall} were saved. However, the fine channelisation ($\sim$97.66~kHz channel bandwidth) alone was sufficient to reduce the intra-channel dispersive smearing to just above one time sample at 550~MHz and below half a sample at 750~MHz. On 2023-12-05, we saved post-correlation mode data instead of the CD data. The signals from 16--17 GMRT antennas were combined in phase, allowing for high sensitivity. Table~\ref{tab:observations} summarises the properties of our observational data.

After obtaining the raw data, we used the \texttt{rficlean} software \citep{2021Maan} to convert them to \texttt{SIGPROC} filterbank files \citep{2011Lorimer} and perform an initial step of radio frequency interference (RFI) excision. \texttt{rficlean} excised periodic (e.g.\ mains power) and impulsive RFI. We then employed the \texttt{DSPSR} software \citep{2011VanStraten} to create dedispersed single-pulse files with 4096 phase bin resolution (187.74~$\mu \text{s}$) based on the best available pulsar ephemeris. Some further manual RFI excision and standard post-processing tasks were performed using the \texttt{PSRCHIVE} software suite \citep{2004Hotan}. We subtracted the profile baselines using the \texttt{PSRSALSA} software \citep{2016Weltevrede}. Additionally, we updated the pulsar's period in the ephemeris and verified it using standard pulsar timing techniques via the \texttt{tempo2} software \citep{2006Hobbs}. The observations were aligned in pulse phase by rotating the 2023-12-05 data by $\sim$0.282 in phase with respect to the 2023-04-24 data, as determined from the timing fit. The data currently lack a robust absolute flux density calibration. For all further data analysis, we used a custom \texttt{Python}-based software suite called \texttt{spanalysis} in version 0.4.7 that we developed for single-pulse analysis tasks.

Throughout this work, we define the phase-resolved profile S/N as the baseline-subtracted profile amplitude per phase bin divided by a robust estimate of the standard deviation in the off-pulse window computed via the interquartile range of the data. We drop an offset of unity for convenience, as is customary. We follow standard pulsar conventions for the total single-pulse profile S/N and compute it via the boxcar equivalent pulse width \citep{2012Lorimer}.

\section{Results}
\label{sec:results}

\begin{table}
\caption{Overview of PSR~B1822$-$09's phenomena analysed in this work with a selection of representative references from the literature.}
\label{tab:overview}
\begin{tabular}{lcr}
\hline\hline
Phenomenon                  & Sections                  & Literature\\
\hline
Profile morphology          & \ref{sec:pulsestacks}     & 1, 2, 3\\
Phase-resolved modulation   & \ref{sec:pulsestacks}     & 4\\
Mode switching              & \ref{sec:modeswitching}, \ref{sec:bfmodediscussion}   & 5, 6, 7\\
Number of modes             & \ref{sec:numberofmodes}, \ref{sec:modeprofiles}, \ref{sec:bfmodediscussion}   & --\\
Fluctuation spectra         & \ref{sec:fluctuationanalysis}     & 4, 6\\
Phase angle spectra         & \ref{sec:fluctuationanalysis}     & --\\
Q-mode modulation           & \ref{sec:fluctuationanalysis}, \ref{sec:fluctuationpropertiesdiscussion}  & 5, 1, 8, 3\\
Q-mode phase-locking        & \ref{sec:fluctuationanalysis}, \ref{sec:qmodephaselockingdiscussion}  & 8, 6, 3\\
Interpole communication     & \ref{sec:correlation}, \ref{sec:qmodephaselockingdiscussion}  & 9, 10\\
Quasi-periodic microstructure   & \ref{sec:microstructure}, \ref{sec:microstructurediscussion}    & 1\\
Mode mixing                 & \ref{sec:modemixing}, \ref{sec:modemixingdiscussion}    & 6\\
Low-intensity pulses        & \ref{sec:weakpulses}      & --\\
Flaring emission            & \ref{sec:flaring}         & --\\
Pulse-energy distributions  & \ref{sec:flaring}         & --\\
Pulse intensity fading      & \ref{sec:pulsefading}     & --\\
Emission geometry           & \ref{sec:emissiongeometrydiscussion}      & 10, 11, 7\\
\hline
\end{tabular}
\tablebib{
(1)~\citet{1994Gil}; (2)~\citet{2016Pilia}; (3)~\citet{2019Yan}; (4)~\citet{2006Weltevrede}; (5)~\citet{1981Fowler}; (6)~\citet{2012Latham}; (7)~\citet{2017Hermsen}; (8)~\citet{2010Backus}; (9)~\citet{1982Fowler}; (10)~\citet{2005DyksA}; (11)~\citet{2007Weltevrede}.
}
\end{table}

We analyse several of PSR~B1822$-$09's emission phenomena in this work. We give an overview of them in Table~\ref{tab:overview} with their section numbers in the paper and a selection of representative references from the literature.

%
%
\subsection{Total integrated pulse profile, profile components, and single-pulse stacks}
\label{sec:pulsestacks}

\begin{figure}
  \centering
  \includegraphics[width=\columnwidth]{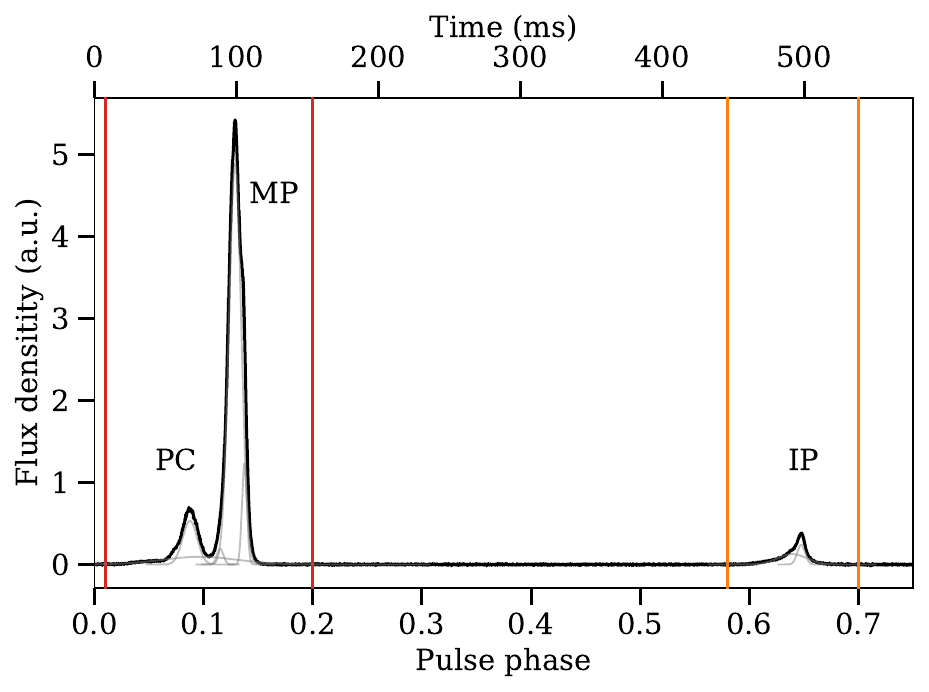}
  \caption{Total integrated pulse profile of PSR~B1822$-$09 formed from our 2023-04-24 data that shows its PC, MP, and IP profile components. The IP is separated by about 0.52 in phase from the MP. The coloured vertical solid lines delineate the two on-pulse phase ranges, and the grey solid lines show our best decomposition into seven von Mises distribution sub-components.}
 \label{fig:pulseprofile}
\end{figure}

\begin{figure*}
  \centering
  \includegraphics[width=\columnwidth]{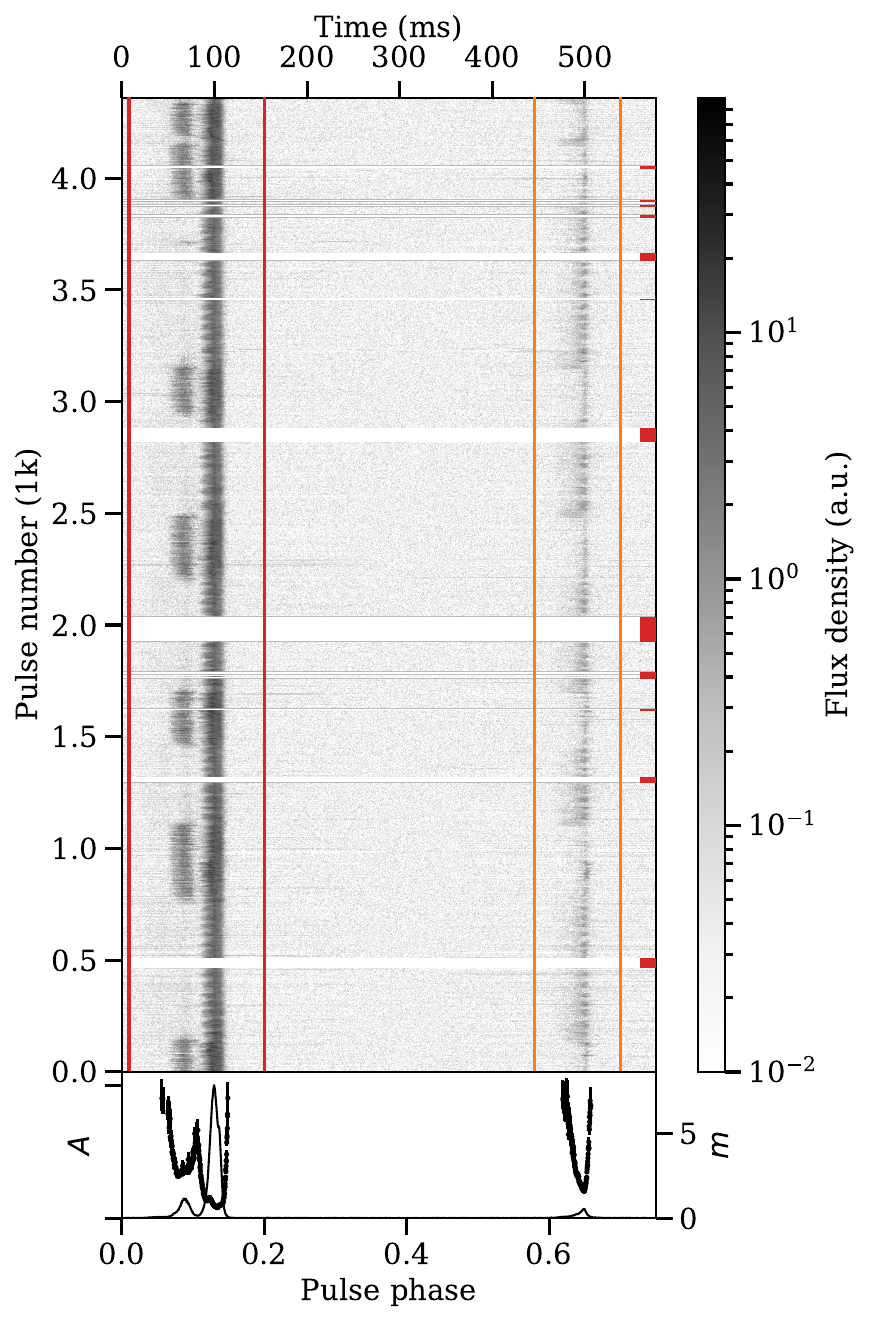}
  \includegraphics[width=\columnwidth]{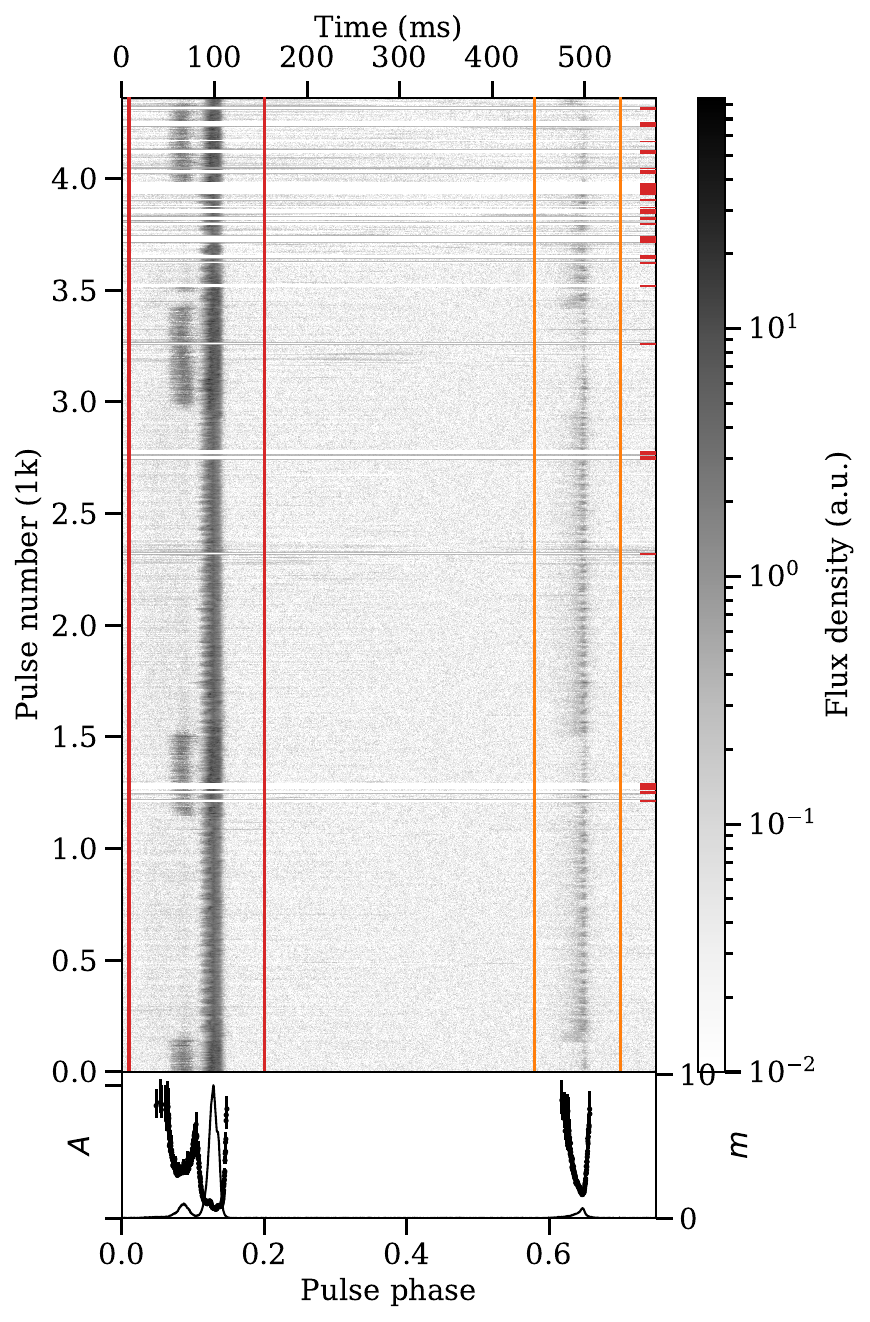}
  \caption{Single-pulse stacks of PSR~B1822$-$09 from our data taken on 2023-04-24 showing several mode transitions. Each stack contains 56~min of data with a gap of 6~min between them for the phase calibrator scan. We applied a logarithmic greyscale mapping to bring out the faint emission in the IP and around the mode transitions. The vertical coloured lines delimit the two on-pulse phase regions, and the horizontal red lines on the right mark the rotations that were excised. The bottom panels show the normalised mean profile amplitudes $A$ (black solid lines; left scale) and the phase-resolved modulation index $m$ (black error bars; right scale).}
 \label{fig:singlepulsestacks01}
\end{figure*}

\begin{figure*}
  \centering
  \includegraphics[width=\columnwidth]{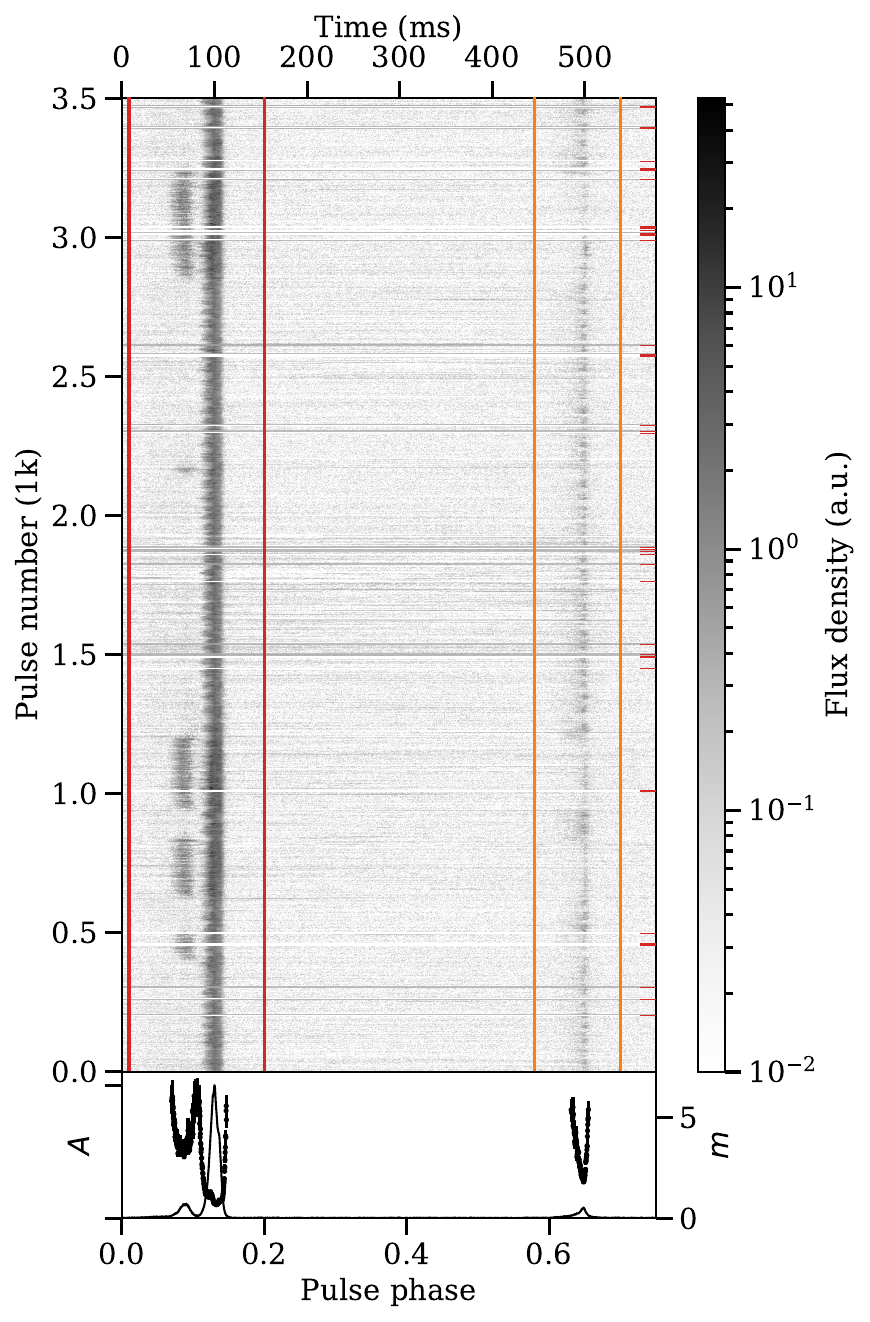}
  \includegraphics[width=\columnwidth]{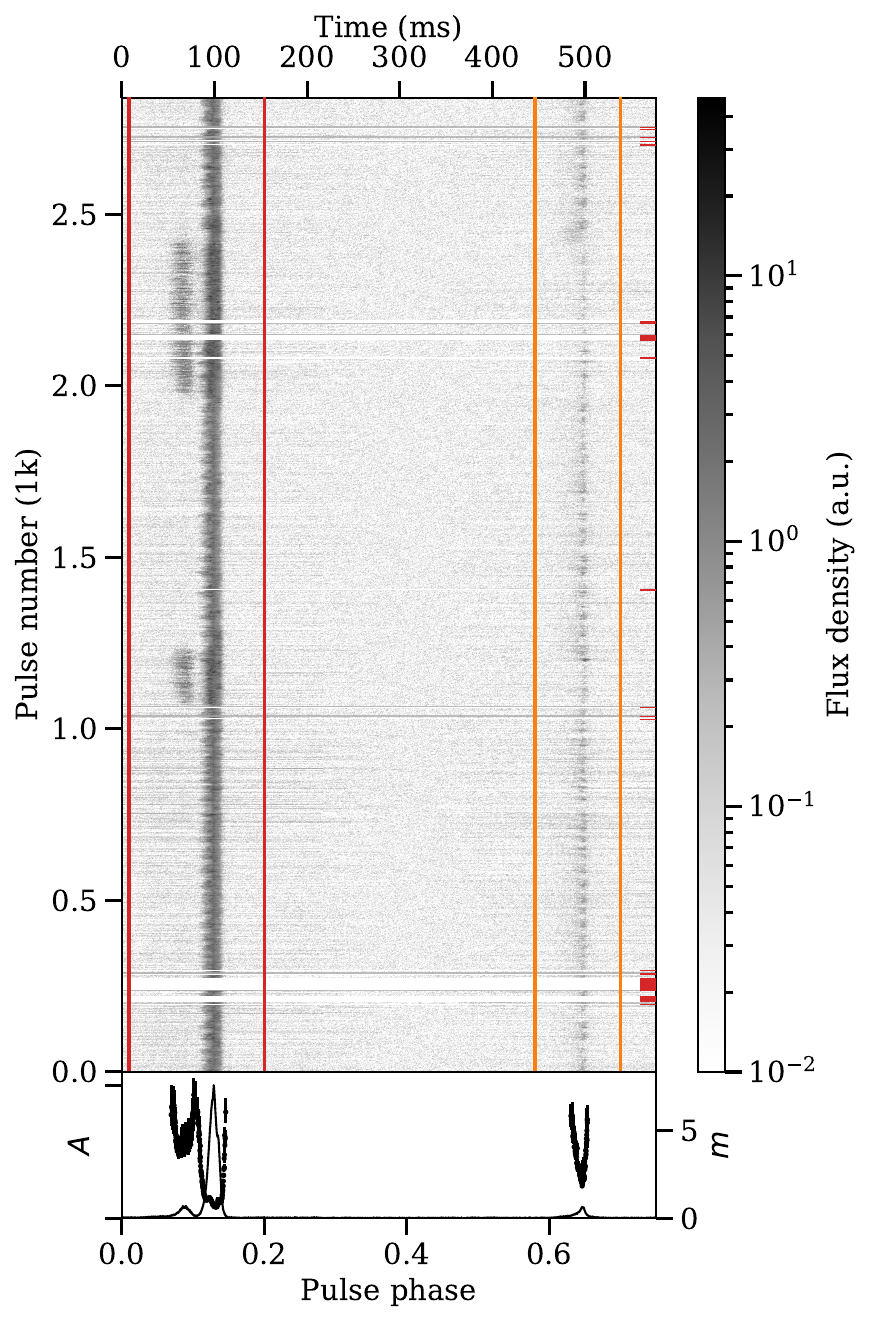}
  \caption{Same as Fig.~\ref{fig:singlepulsestacks01}, but for the uGMRT data obtained on 2023-12-05.}
 \label{fig:singlepulsestacks02}
\end{figure*}

Figure~\ref{fig:pulseprofile} shows the pulsar's total integrated pulse profile formed from the 2023-04-24 data with the different profile features clearly visible. Its integrated profile consists of a main pulse and interpulse on-pulse phase regions, delineated by the vertical red and orange lines that we placed visually. The main pulse region contains the precursor (PC) and the main pulse (MP) profile components\footnote{They were called components 1 and 2 (C1 and C2) in earlier work \citep{1981Fowler, 1994Gil}.}. The interpulse (IP) is separated from the MP by about half a rotation of the star ($\sim$0.52 phase peak-to-peak), which likely makes it a genuine IP, that is, emission from the polar cap on the opposite side of the star. The grey solid lines in Fig.~\ref{fig:pulseprofile} show our best profile decomposition into scaled von Mises distribution sub-components:
\begin{equation}
    f(\varphi | \mu, \kappa) = A \exp \left\{ \kappa \left[ \cos \left( \varphi - \mu \right) - 1 \right] \right\},
    \label{eq:vonmisesdistribution}
\end{equation}
where $\varphi \in [0, 2 \pi]$ is the pulsar rotational phase, $A$ the amplitude, $\mu$ the mean, and $\kappa$ the concentration, that is, the inverse of the Gaussian variance. The definition above is slightly adjusted from its standard form for numerical convenience and compatibility with the implementation in \texttt{PSRCHIVE}. In this model, the profile consists of seven sub-components: two in the PC, four in the MP, and two in the IP, where one component spans both the PC and MP. In particular, the PC consists of a bright primary sub-component in the centre and a significantly fainter and wider plateau component, which also contributes to the bridge emission and the MP. The MP is complex, with a strong primary sub-component on the leading side, a somewhat fainter trailing shoulder component, and a tiny bump at the leading edge. The IP comprises a narrow, bright component at the trailing edge and a wider and fainter component at the leading side responsible for the IP's asymmetry. The PC has roughly 13\% of the peak flux of the MP, and the IP 7\%. Integrating across the profile components, we find that the MP carries about 79\% of the total pulse-averaged flux density. The PC and IP contribute 14 and 7\% to the mode-averaged total. Interestingly, the PC is almost exactly twice as bright as the IP in both the peak and pulse-averaged sense in this frequency band, which could point to a common origin, as was suggested earlier \citep{2005DyksA}.

We present the single-pulse stacks for the 2023-04-24 observation in Fig.~\ref{fig:singlepulsestacks01}. The left panel shows the first 56~min of data and the right panel the second, with a gap of 6~min in between. Figure~\ref{fig:singlepulsestacks02} shows the same but for the observation on 2023-12-05. In both figures, we zoomed in slightly in pulse phase for clarity. The bottom panels show the normalised mean pulse profile computed over that pulse stack as black solid lines referenced to the left amplitude $A$ scale and the phase-resolved time-domain modulation index
\begin{equation}
    m_i = \frac{ \sigma_i }{ \mu_i }
    \label{eq:modulationindex}
\end{equation}
with black error bars referenced to the right modulation $m$ scale, where $\sigma_i$ is the standard deviation and $\mu_i$ is the mean Stokes I profile intensity at the phase bin $i$. We show the modulation index only for phase bins where the mean profile exceeds 1--2\% of its peak amplitude and estimated its uncertainties robustly by bootstrap resampling the data 100 times. The on-and-off switching of the PC component is clearly visible. The switching time is significantly shorter (more frequent) in the first 2023-04-24 scan compared with the second. The pulsar switched into its B-mode nine times during that observation. The switching is slightly less frequent in the 2023-12-05 data, where the pulsar was in the B-mode seven times. The phase-resolved modulation index is qualitatively similar in each of the four scans. The most stable emission (lowest $m_i$) originates near the centre of the MP, approximately aligned with the hump in the profile. The MP's leading edge has a slightly higher modulation with a small S-shaped bump visible in the $m_i$ curve. The modulation index increases across the bridge between the MP and PC until it decreases again in the PC phase range, at which point it reaches another local minimum. The modulation index there is several times higher than in the MP and is dominated by the mode switching process. The modulation index increases roughly symmetrically at the profile edges, as expected. The IP shows a V-like $m_i$ curve that is practically featureless but slightly asymmetric towards its leading profile edge, tracing its profile shape. The minimum occurs near the IP's profile peak.

Our measurements of PSR~B1822$-$09's pulse profile and $m_i$ curve at 650~MHz look qualitatively very similar to what was reported in the literature. In particular, compared with the measurements obtained at 1.4~GHz by \citet{2006Weltevrede}, the ratio between the PC and MP amplitude is higher at higher radio frequencies. The trailing hump in the MP is clearly visible at both frequencies. The shape of the $m_i$ curve is very similar, too. It is almost flat across most of the IP in both cases. The characteristic S-shaped bend near the leading flank of the MP is reproduced in almost the same way, and the minimum $m_i$ aligns with the MP profile hump at both frequencies. The rise in $m_i$ towards the MP's trailing edge looks slightly steeper in the 1.4~GHz data, but this could also be due to their limited phase resolution. The absolute $m_i$ values are comparable, but we caution that slight deviations are expected because of the differing observing frequencies and computation methodologies. We analyse the pulsar's switching behaviour in detail in the following sections.

%
%
\subsection{Understanding the pulsar's mode switching}
\label{sec:modeswitching}

The first step in any mode switching pulsar analysis is to determine how many different stable emission modes a pulsar exhibits and to separate them in the pulse stacks. This mode classification or sequencing is a crucial first step for any following analyses and thus greatly influences any conclusions drawn. This is usually done visually from the pulse stacks or by looking at the intensity of emission in distinct pulse phase windows, where in the case of PSR~B1822$-$09 the presence of its PC is taken as the mode determining feature \citep{1981Fowler, 1981Morris, 1994Gil, 2010Backus}. The mode classification using the latter method can be performed quasi-automatically \citep{2017Hermsen, 2019Yan}. However, it assumes that the pulsar emits in a single mode at a given time; that is, it is a binary classification. In contrast, \citet{2012Latham} found evidence of mode mixing in PSR~B1822$-$09 in the form of short emission sequences during which it was apparently emitting simultaneously in both modes. A different classification method is to decompose the single-pulse profiles into their Eigen mode profiles and analyse their mixture weights as a function of pulse number \citep{2024Cao}, which naturally incorporates mode mixing. Here, we describe a new mode classification method that uses time series of profile features as input.

\subsubsection{A hidden Markov switching model for the pulsar's moding process}

\begin{figure}
  \centering
  \includegraphics[width=\columnwidth]{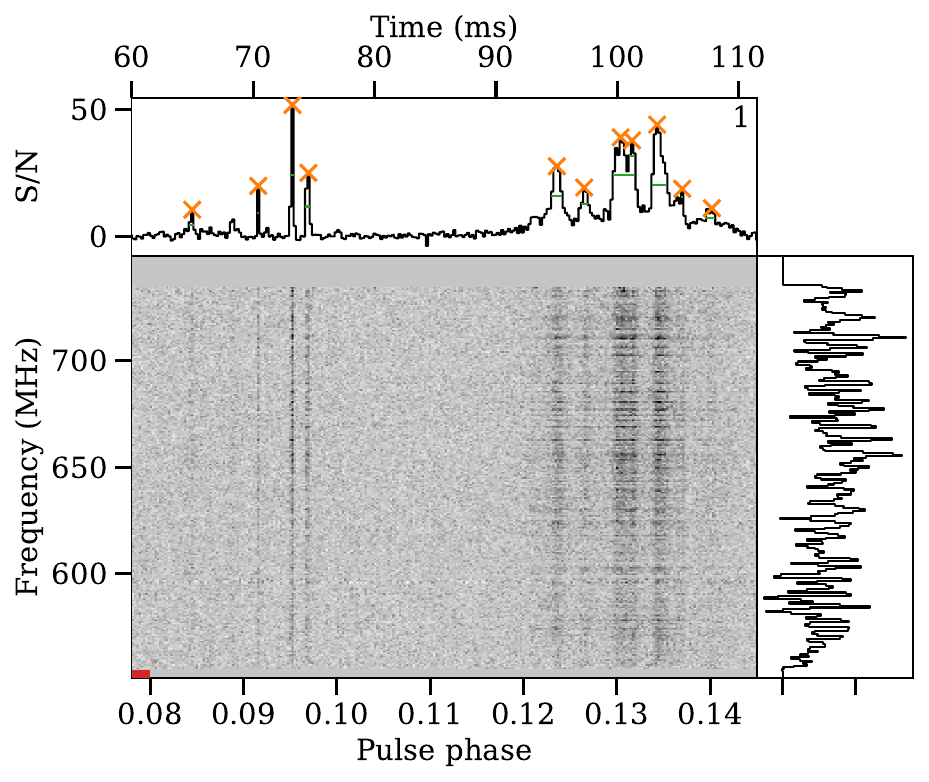}
  \caption{Example of a single-pulse dynamic spectrum and profile with the significantly peaked profile components and their full widths at half prominence highlighted. The right panel shows the mean power spectral density or radio spectrum in the on-pulse region. A sub-threshold peak is visible in the PC phase window on the left.}
 \label{fig:profilefeatures}
\end{figure}

\begin{figure*}
  \centering
  \includegraphics[width=\textwidth]{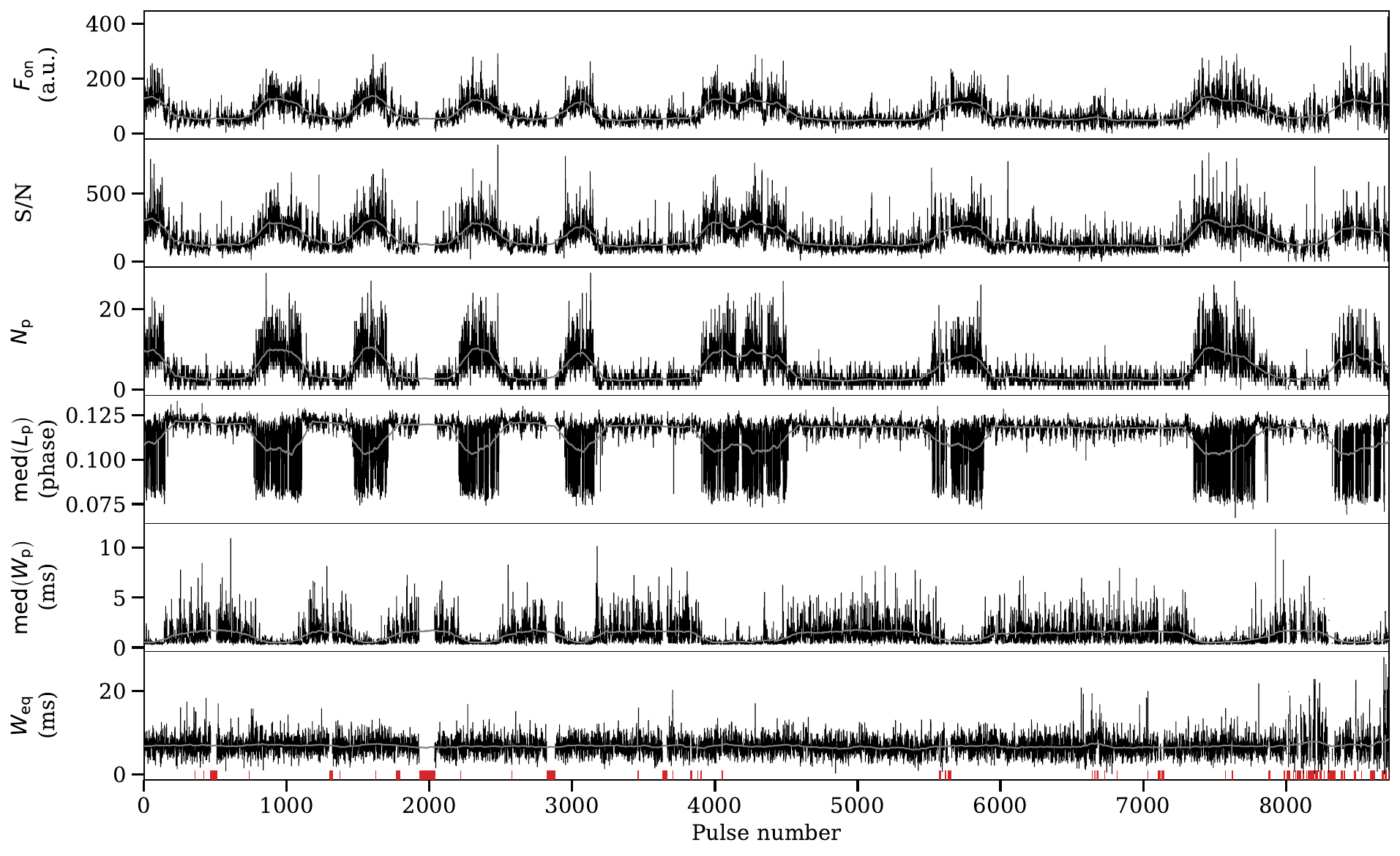}
  \caption{Timeline showing several measured features from the entire 2023-04-24 single-pulse stack. Specifically, we show time series of the on-pulse fluence $F_\text{on}$ in the combined PC and MP phase range, the S/N, the number of peaks $N_\text{p}$ in the profile, their median locations $L_\text{p}$ in pulse phase, their full widths at half prominence $W_\text{p}$, and the boxcar equivalent pulse width $W_\text{eq}$. The grey lines depict running mean smoothings of the data using a 150-rotation wide sliding window. The red vertical markers highlight rotations that were RFI excised.}
 \label{fig:featuretimeline}
\end{figure*}

With the aim of modelling the pulsar's mode switching, we measured several features from the cleaned single-pulse stacks shown in Figs.~\ref{fig:singlepulsestacks01}--\ref{fig:singlepulsestacks02}. In particular, we measured time series of various observable quantities against pulse number, that is, a particular quantity measured for each single pulse. We call those `rotation time series'. Among others, we measured rotation time series of the on-pulse fluence $F_\text{on}$, S/N, and boxcar equivalent pulse width $W_\text{eq}$. Additionally, we employed a peak-finding algorithm to determine the number of prominent S/N $\geq 10$ pulse profile components or peaks $N_\text{p}$ in the on-pulse region, their locations $L_\text{p}$ in phase, and their full widths at half prominence $W_\text{p}$. We also required each peak to have a S/N $\geq 5$ prominence with respect to its surroundings. Prominence or relative height measures the independence of a (topographic) peak, that is, our choice ensures that each identified peak is sufficiently independent from the other nearby peaks. In Fig.~\ref{fig:profilefeatures}, we show an example single pulse with the peaks and their full widths highlighted. The orange crosses mark the peaks that fulfil the above criteria and are included in the $N_\text{p}$ count. Because of the S/N threshold, $N_\text{p}$ is a robust lower limit for the actual number of peaked components in the profile. In particular, we confirmed that an additional small number of peaks were visible in the data close to the baseline noise limit, as is shown in Fig.~\ref{fig:profilefeatures}. For each pulsar rotation, we computed the median over all peak locations and widths as a summary statistic, which we denote as $\text{med} (L_\text{p})$ and $\text{med} (W_\text{p})$, respectively. We considered the combined PC and MP on-pulse phase range for this analysis, while the IP range was excluded. Figure~\ref{fig:featuretimeline} shows a comparison of the various feature time series. The mode switching is clearly visible in the fluence, S/N, $N_\text{p}$, $L_\text{p}$, and $W_\text{p}$ series. The $W_\text{eq}$ times series is too noisy at the single-pulse resolution. The fluence and S/N series beautifully reflect the doubling in pulse-averaged flux density when the pulsar switches from its Q to its B-mode. The $N_\text{p}$ and $W_\text{p}$ time series in Fig.~\ref{fig:featuretimeline} nicely show the switching between single pulses with a handful of peaked components of large width (Q-mode) and pulse profiles with 10--25 almost featureless shot-like micropulse components (microstructure spikes) of extremely narrow width around one phase bin (B-mode). In other words, the $N_\text{p}$ and $W_\text{p}$ time series show alternating behaviour. The $L_\text{p}$ time series nicely reflects the negative shift in median peak location due to the appearance of the PC profile component and its many peaks.

In a second step, we employed a hidden Markov model to describe the pulsar's mode switching process. It consists of two parts: (1) a model for the hidden transition process of the pulsar and (2) a model for the observed feature timelines for each hidden state. In particular, we modelled the pulsar as a stochastic system that exhibits $N$ stable emission states $S_t = \{ 0, 1, \ldots, N - 1 \}$. Those states are hidden or latent, that is, not directly observable. We assumed that they obey the Markov property so that a given state $S_t$ only depends on the state $S_{t-1}$ directly before it. This way, a first-order Markov chain can describe the switching process. The state transitions are represented using the $N \times N$ square transition matrix:
\begin{equation}
    P \left( S_t | S_{t-1} \right) = \left[ p_{ij} \right] =
    \begin{bmatrix}
    p_{00} & p_{10} & \ldots & p_{N-1 0}\\
    p_{01} & p_{11} & \ldots & p_{N-1 1}\\
    \vdots & \vdots & \ddots & \vdots\\
    p_{0 N-1} & p_{1 N-1} & \ldots & p_{N-1 N-1}\\
    \end{bmatrix}.
    \label{eq:transitionmatrix}
\end{equation}
The matrix describes the transition probability from state $S_j$ (column) into the state $S_i$ (row); that is, it is read by looking at the column $j$ first and the row $i$ second. The off-diagonal elements $p_{ij}, i \neq j$ describe the transition probabilities between states, and the diagonal elements $p_{ii}$ describe how stable a given state is in time (self-transition), where $i, j \in \{ 0, 1, \ldots, N - 1 \}$. The matrix is left stochastic, meaning that all columns must sum to unity, $\sum_{j=0}^{N-1} p_{ij} = 1, \: \forall i$. We then combined the hidden Markov model with a model for the observed feature time series for each stable pulsar emission state. Namely, we assumed that each state produced a measured time series $y_t$ with a given mean $\mu_{S_t}$, a stochastic normally distributed white-noise contribution $\epsilon_{S_t} \sim N(0, \sigma_{S_t}^2)$, where $\sigma_{S_t}^2$ is the Gaussian variance, and further autoregressive terms that depend on the previous values $y_{t-i}$ of the measured time series itself at the time lags $i$. The latter is called autoregressive behaviour of a time series and is analogous to recursion. Formally, this is given by
\begin{equation}
    y_t = \mu_{S_t} + \epsilon_{S_t} + \sum_{i = 1}^{p} \phi_i \: \left( y_{t - i} - \mu_{S_{t-1}} \right),
    \label{eq:switchingmodel}
\end{equation}
where $p$ is the order of the autoregression and $\left| \phi_i \right| < 1$ are the autoregression coefficients. The term $\phi_i$ encodes the transition behaviour of the observed time series. While each hidden state $S_t$ only depends on the previous $S_{t-1}$, the observed time series generated by the hidden states depends on the previously observed values of itself up to a given time lag $p$. More concretely, $y_t$ can model decaying or oscillatory behaviour depending on the choice of $\phi_i$. Including the autoregressive terms in the model is important, as the measured feature time series show clear autoregressive behaviour. We verified this by computing their (partial) autocorrelation functions. For instance, the partial autocorrelation within the $F_\text{on}$ and $N_\text{p}$ time series decreases roughly exponentially from zero lag and stays significant ($> 0.021$) until about lags 20--30 with a few non-negligible peaks beyond that. However, fitting an autoregressive switching model of that order is computationally challenging. Therefore, we limited ourselves to autoregressive orders $p = 1 - 2$ and a global non-switching white noise term $\epsilon_t$. This choice limits the capability of our model to describe the data only slightly while providing drastic runtime savings. It only affects the modelling of the transition behaviour between states, not their identification or total number. That is because the lowest-order coefficients $\phi_i$ are generally the largest and contribute the most to the autoregression. The combined model is called a hidden Markov switching autoregression model, originally developed for financial time series data in econometrics \citep{1989Hamilton, 2020Hamilton}. It is similar to a general mixture model combined with a temporal transition framework.

We then systematically used each measured feature time series described above (Fig.~\ref{fig:featuretimeline}) as input for our Markov switching model and estimated its free parameters using an iterative maximum likelihood regression procedure. While we experimented with several different methods, our current implementation uses the software packages \texttt{SciPy} \citep{2020Scipy} and \texttt{statsmodels} \citep{2010Statsmodels}. Based on the best-fitting parameters, we computed the marginalised probability time series $p(S_t)$ of the pulsar being in state $S_t$ at a given rotation, the average durations of each state, and the overall state fractions. The $N_\text{p}$ time series had the highest mode-separation power at the single-pulse level out of the features we tested. That is because the number of peaked profile components is little affected by and therefore robust to the overall single-pulse brightness modulation above a limiting threshold peak S/N, which we set reasonably high (10 S/N). More importantly, the fact that the PC profile has a large number of peaked components makes $N_\text{p}$ a reliable indicator for its activity. A less sensitive instrument will detect fewer peaked components, predominantly in the PC. The classification stays robust as long as the $N_\text{p}$ distributions differ sufficiently between the modes. The above is valid at the single-pulse level. However, the picture is likely different when rotation-integrated profiles are used for the mode classification. For instance, some previous publications used rotation-integrated data (e.g.~ten rotations or 10-s integrations) for the mode classification, which are insensitive to rapid switching and less sensitive to profile structure due to the averaging process (e.g.~partially \citealt{1994Gil}, \citealt{2017Hermsen}).

\subsubsection{Testing the number of stable emission modes exhibited by the pulsar}
\label{sec:numberofmodes}

\begin{figure*}
  \centering
  \includegraphics[width=\textwidth]{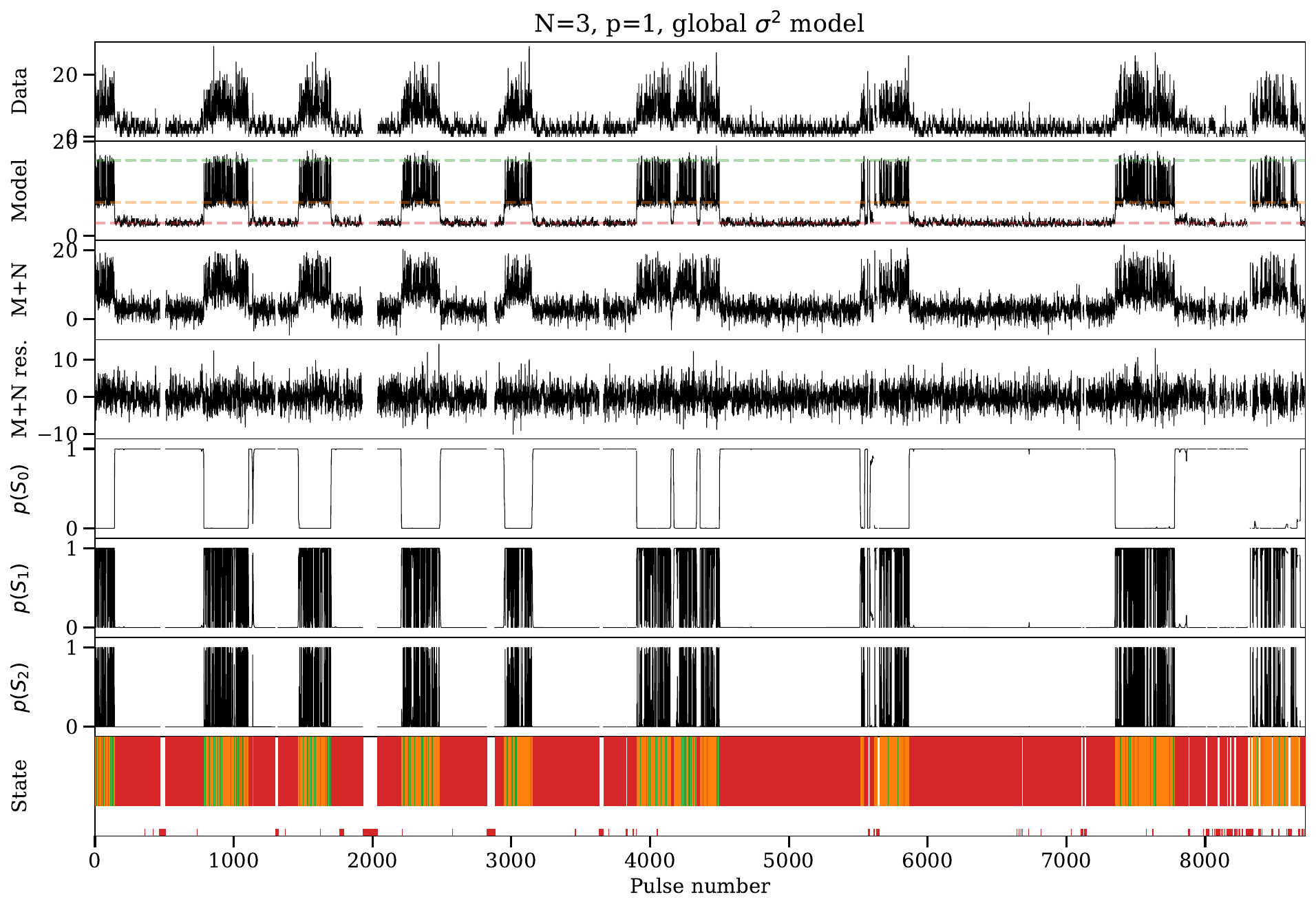}
  \caption{Results from fitting our Markov switching models to our measured $N_\text{p}$ time series data from the entire 2023-04-24 single-pulse stack. We show the best fit of our three-state model with first-order autoregression and a globally constant variance. The panels show, from top to bottom, the $N_\text{p}$ time series data, the best-fitting noise-free switching model, the model with added Gaussian white noise with the estimated variance (M+N), the residuals of that (data - M+N), the state probability time series $p(S_t)$, and the most likely emission state for each rotation coloured separately. Red corresponds to $S_0$, orange to $S_1$, and green to $S_2$. We show the best-fitting state means with dashed lines of the same colour in the model panel. As above, the red vertical markers highlight rotations that were RFI excised. The model describes the data well, with almost white residuals.}
 \label{fig:markovmodelfitresults}
\end{figure*}

\begin{figure}
  \centering
  \includegraphics[width=\columnwidth]{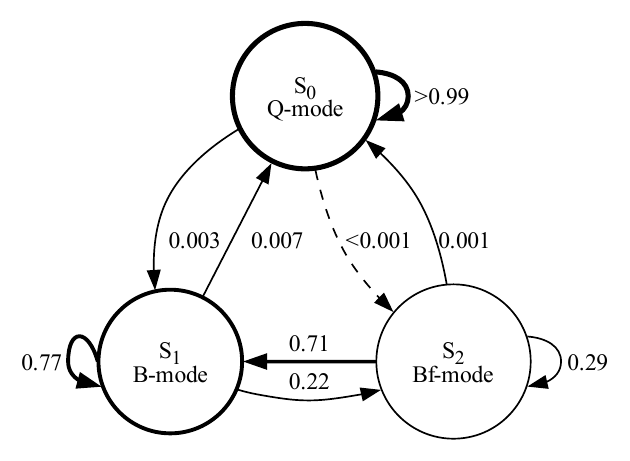}
  \caption{Visualisation of the state transition probabilities of our best-fitting three-state Markov switching model. The circles represent the states, and the arrows represent the state transitions. The line thickness is roughly proportional to the transition probability, with low-probability transitions drawn using dashed lines. The identified states are the quiescent Q-mode, the bright B-mode, and a newly identified flaring Bf-mode.}
 \label{fig:markovstatediagram}
\end{figure}

\begin{figure}
  \centering
  \includegraphics[width=\columnwidth]{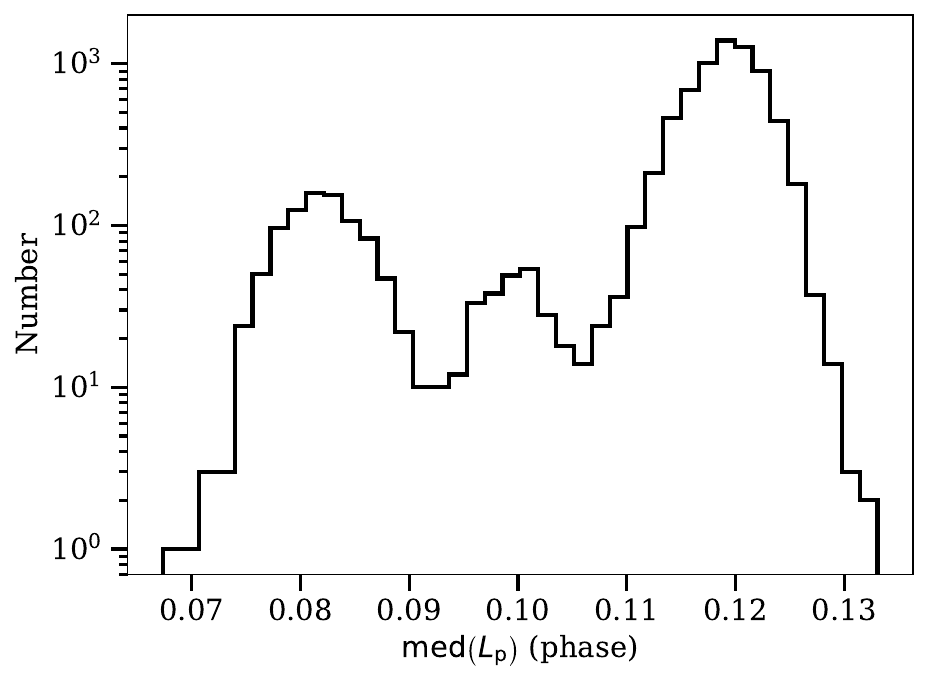}
  \includegraphics[width=\columnwidth]{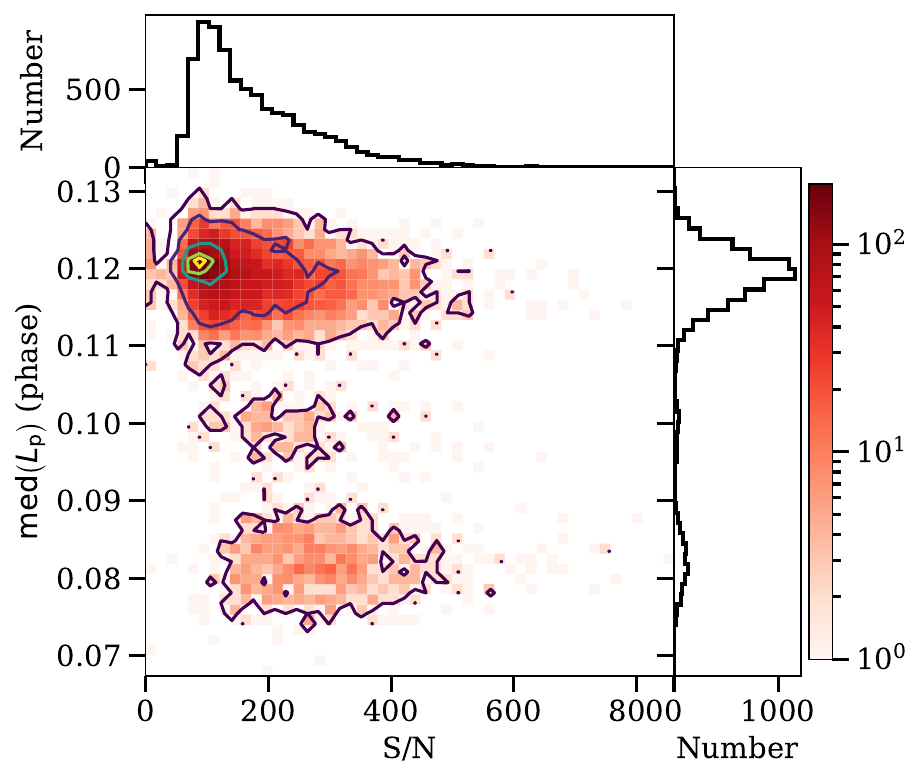}
  \includegraphics[width=\columnwidth]{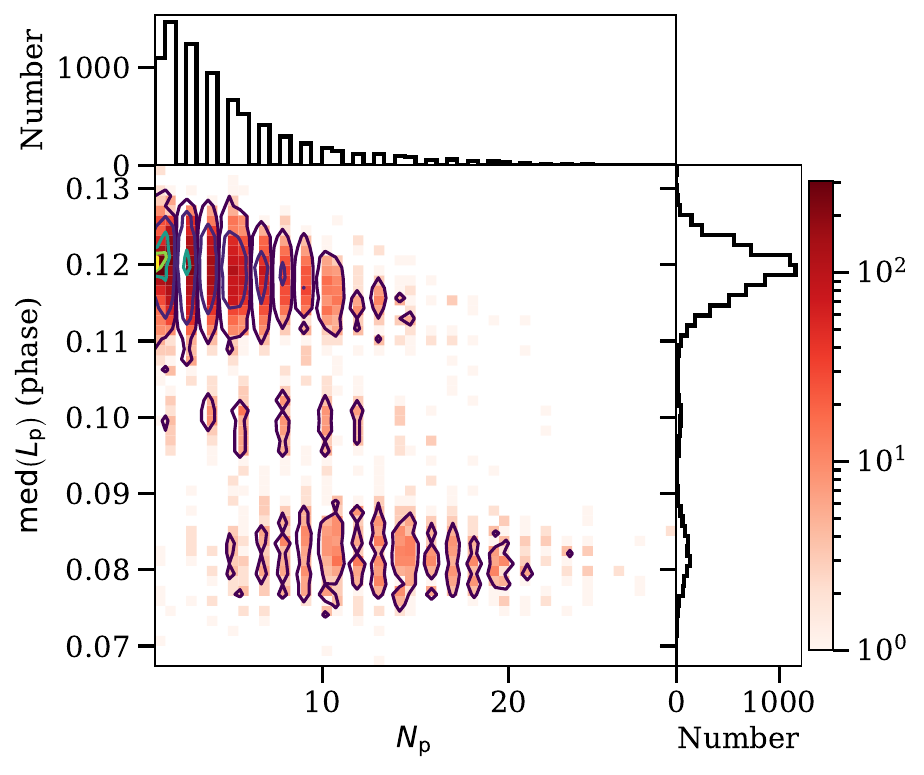}
  \caption{Visual analysis of the number of emission modes present in the data. Top: 1D histogram of the median location of the peaked profile components shown in logarithmic scale. Middle and bottom: Two profile features plotted against each other in 2D (logarithmic scale) and 1D histograms (linear scale) on the sides. We show contours at 1, 10, 50, 75, and 90\% of the maximum bin count. The histograms indicate the presence of at least three stable emission modes in the pulsar and independently confirm our conclusions from the Markov model analysis.}
 \label{fig:featurespace}
\end{figure}

Using the Markov switching model described above, we investigated several questions. Firstly, we were interested in how many stable emission modes the pulsar exhibits. Apart from the two well-known ones (Q and B-mode) that are straightforward to identify visually, there could be additional, more subtle modes hidden in the data. Alternatively, there could be mode mixing at certain times, as reported by \citet{2012Latham}. We evaluated this question by successively increasing the number of states $N$ of the Markov switching model, fitting it to the measured $N_\text{p}$ feature time series data, and comparing the resulting residual plots and values of the Akaike information criterion (AIC; \citealt{1974Akaike}) to prevent over-fitting. Specifically, we iteratively tested Markov switching models with $N = 2 - 7$ states, a first-order autoregressive term ($p = 1$), and a globally constant (non-switching) white noise contribution (Eq.~\ref{eq:switchingmodel}). A two-state model is strongly disfavoured by the data in comparison with the other models tested. For all further analysis, we selected the three-state model as the lowest-complexity model that describes the data well, with almost white residuals, as we demonstrate below. Figure~\ref{fig:markovmodelfitresults} shows the diagnostic plot of our best-fitting three-state model. The panels present from top to bottom: the $N_\text{p}$ time series data, our best-fitting noise-free model, the model with additive Gaussian noise with the given best-determined global variance (M+N), the residual of the model plus noise (data - M+N), and the marginalised probability time series $p(S_t)$ for each state. The bottom panel of Fig.~\ref{fig:markovmodelfitresults} visualises the most likely mode that was active during each pulsar rotation, where we used a different colour for each mode. Red corresponds to $S_0$, orange to $S_1$, and green to $S_2$. While some minor structure is left in the M+N residuals shown in Fig.~\ref{fig:markovmodelfitresults} panel 4, the residuals appear almost white, indicating a good agreement between the model and data.

The best-fitting model parameters are $\mu_0 = 2.68 \pm 0.04$, $\mu_1 = 6.98 \pm 0.07$, $\mu_2 = 15.8 \pm 0.1$, $\sigma^2 = 3.75 \pm 0.06$, and $\phi_1 = 0.27 \pm 0.01$. The fact that $\phi_1$ deviates significantly from zero means that the time series shows clear autoregressive behaviour. The reduced $\chi^2$ of the fit is 3.5 for 11 free model parameters and 8096 degrees of freedom. The best-fitting state transition matrix is
\begin{equation}
    P \left( S_t | S_{t-1} \right) =
    \begin{bmatrix}
    $>0.99$	    &	0.007	&	0.001\\
    0.003       &	0.77	&	0.71\\
    $<0.001$	&	0.22	&	0.29\\
    \end{bmatrix}.
\end{equation}
The corresponding average state durations were 371.7, 4.2, and 1.3 rotations, or 285.9, 3.2, and 1~s, for $S_0$ through $S_2$. The pulsar spent approximately 66, 26, and 8\% of the time in the different states. Figure~\ref{fig:markovstatediagram} visualises the transition probabilities between the modes in a Markov state diagram. $S_0$ is a long-lived and stable dominant mode that is present almost 70\% of the time. The number of single-pulse profile peaks $N_\text{p}$ is small ($\sim$3). There is a small probability of transitioning from $S_0$ to $S_1$, whereas a direct transition from $S_0$ to $S_2$ is highly suppressed. $S_1$ is relatively stable and is present in our data for about 25\%. The typical number of peaks is higher ($\sim$7). The transition probability to $S_2$ is considerable at around 20\%. $S_2$ is an extremely transitory and short-lived mode with a large number of profile peaks ($\sim$16) present, which decays to $S_1$ with around 70\% probability. It is active in our data for about 8\% of the time.

We  ran our model on the 2023-12-05 data separately, which are shorter and slightly less sensitive. The best-fitting parameters are quite similar with $\mu_0 = 1.58 \pm 0.03$, $\mu_1 = 4.60 \pm 0.07$, $\mu_2 = 12.1 \pm 0.1$, $\sigma^2 = 2.54 \pm 0.05$, and $\phi_1 = 0.21 \pm 0.02$. The reduced $\chi^2$ of the fit was 2.3 for 6013 degrees of freedom. The average state durations were 301.4, 5.1, and 1.4 rotations, and the state occupancy was roughly 77, 19, and 5\%. The absolute $N_\text{p}$ decreased slightly, as expected, given the lower sensitivity of the observation. However, the mode identification remained the same. Combining both observations, the mean values weighted by the effective number of observed pulses are 341.7, 4.6, and 1.3 rotations for the average state durations and 71, 23, and 7\% for the state occupancies.

Our results agree well with the literature. For instance, \citet{2017Hermsen} found that the pulsar spent about 64\% of its time in the Q-mode, which almost exactly matches our result for state $S_0$ in the 2023-04-24 analysis and is close to our overall value. The average Q-mode duration in their data was 270~s, or about 347 rotations. This is very close to our overall average value ($\sim$342 rotations). Any difference might arise because \citet{2017Hermsen} used rotation-averaged data in their analysis. These suffer from misalignment issues between integration boundaries and mode switches and are not (or less) sensitive to mode mixing or pulse intensity fading near mode switches, as we discuss in Sect.~\ref{sec:spmorphology}. RFI also plays a more important role, affecting entire 10-s integrations. For the same reason, the \citet{2017Hermsen} analysis was insensitive to rapid switching, as we see for states $S_1$ and $S_2$. This and the fact that we performed a multi-mode rather than two-mode classification prevents a direct comparison of the B-mode measurements. Similarly, our measurements agree with those reported by \citet{2012Latham}.

An independent way to investigate the number of emission modes present and verify our conclusions from the Markov switching model analysis is to look at histograms of the feature time series and try to decompose them into disjoint components belonging to each mode. We show examples of this in Fig.~\ref{fig:featurespace}. The top panel shows a 1D histogram of the median peak location med($L_\text{p}$) determined using the peak-finding algorithm described above on the single-pulse profiles (Fig.~\ref{fig:profilefeatures}). The middle and bottom panels show scatter plots of two profile features against each other in what we call a `feature space', with their 1D histograms at the sides. In the middle panel, we show med($L_\text{p}$) plotted against the total single-pulse S/N, and in the bottom panel, we show med($L_\text{p}$) against the number of peaks $N_\text{p}$. In each case, the data reveal three obvious clusters in the med($L_\text{p}$) domain, which corresponds to single-pulse profiles where the peaked profile components are dominant either in the PC (near 0.08 phase) or the MP (near 0.12 phase). The middle peak comes from single-pulses with equal numbers of peaked profile components in the PC and MP phase ranges, which we denote as `equalised pulses'. Bridge emission with the MP inactive was not observed. More subtle clusters exist in the S/N domain, where the MP-dominated pulses have a most likely S/N around 100 (mode and contour), while the PC-dominated pulses cluster around 300 S/N. The equalised pulses occupy the S/N range in between, near 200 S/N. The bottom panel of Fig.~\ref{fig:featurespace} gives the same picture for the $N_\text{p}$ domain. The MP-dominated pulses cluster around 2--3 peaked profile components, the PC cluster is wide and centred at 12--13 peaks, and the equalised pulses cluster in the intermediate $N_\text{p}$ range centred at 7. Thus, the visual histogram decomposition results in $N_\text{p}$ centroid values that are very similar to the means from our Markov model analysis. In conclusion, the visual inspection of the feature histograms provides additional evidence that the pulsar exhibits at least three emission modes. We describe the profile morphology of each mode in the following section.

\subsubsection{Mode-separated pulse profiles}
\label{sec:modeprofiles}

\begin{figure*}
  \centering
  \includegraphics[width=\columnwidth]{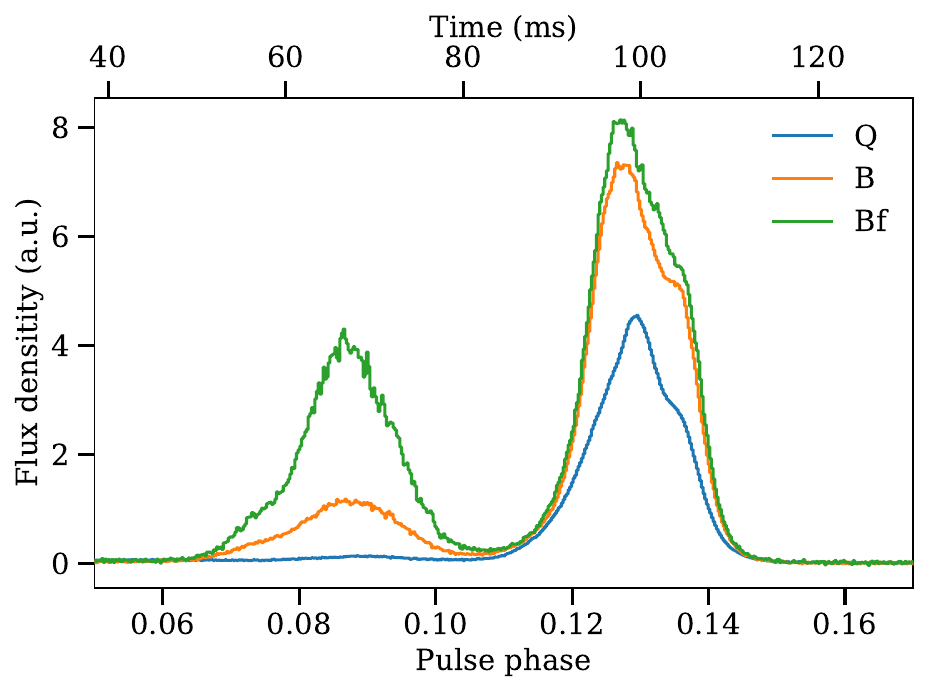}
  \includegraphics[width=\columnwidth]{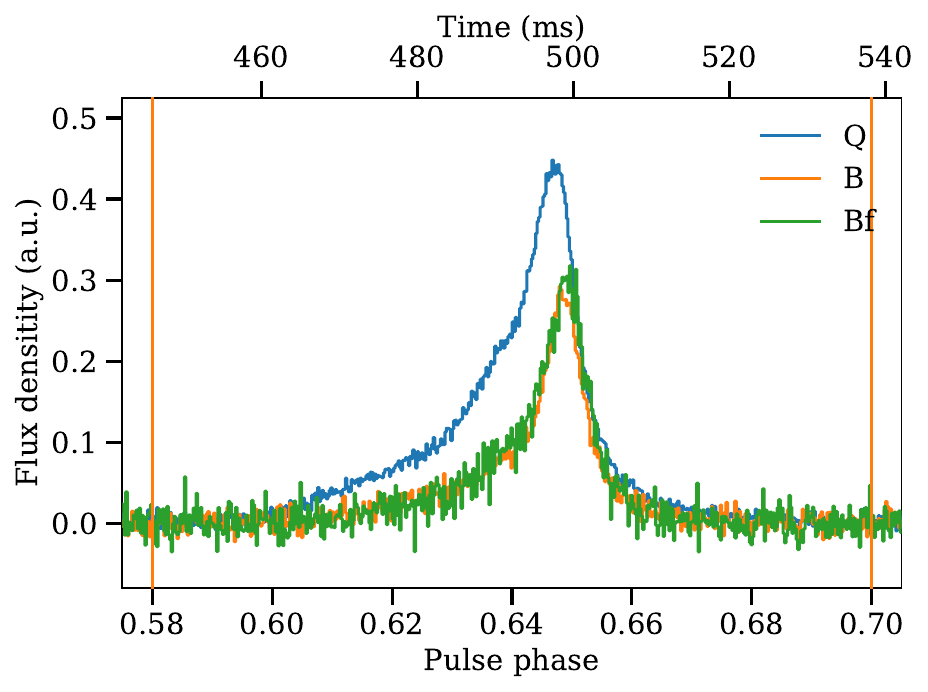}
  \caption{Comparison of the mode-separated mean pulse profiles. The panels show zooms onto the combined PC and MP phase range (left) and the IP phase window (right). The quiescent Q-mode (blue), the bright B-mode (orange), and the bright flaring Bf-mode (green) are plotted on the same absolute scale.}
 \label{fig:modecomparison}
\end{figure*}

\begin{table}
\caption{Component-resolved pulse-averaged flux density contributions for each emission mode.}
\label{tab:modefluxes}
\begin{tabular}{lcccc}
\hline
Mode    & Total     & PC    & MP    & IP\\
        & (a.u.)    & (\%)  & (\%)  & (\%)\\
\hline
Q       & 0.09      & 6     & 81    & 10\\
B       & 0.15      & 17    & 80    & 3\\
Bf      & 0.21      & 34    & 63    & 2\\
\hline
\end{tabular}
\tablefoot{The percentages are affected by rounding.}
\end{table}

Our next aims were to match the states identified by our Markov switching model to the physical emission modes of the pulsar, look at their mode-separated pulse profiles, and characterise them. To do that, we converted the most likely state membership time series shown in Fig.~\ref{fig:markovmodelfitresults} bottom panel to binary selection masks, one for each state. We then applied the masks to the single-pulse stack data, selecting only the rotations that belonged to that particular state. Figure~\ref{fig:modecomparison} compares the resulting mode-separated mean pulse profiles plotted on the same absolute scale. They are the mean profiles computed across the selected subset of single-pulses corresponding to a particular state. For clarity, we show the combined PC and MP (left panel) and the IP phase range (right panel) separately. Emission outside those phase windows was consistent with the baseline noise for all three states. Based on the appearance of their mode-separated pulse profiles, we identified state $S_0$ as the quiescent Q-mode, $S_1$ as the bright B-mode, and $S_2$ as a newly discovered bright flaring mode that we call `Bf-mode'. We describe each mode's characteristics in the following.

Considering the entire on-pulse window (PC, MP, and IP), the pulsar emitted continuously in each rotation unaffected by RFI; that is, there was no evidence for nulling in our data. We show the measured total pulse-averaged flux densities and the component-resolved flux density fractions for each mode in Table~\ref{tab:modefluxes}. Namely, the PC, MP, and IP emission accounted for about 6, 81, and 10\% of the total emission in the Q-mode, roughly 17, 80, and 3\% in the B-mode, and approximately 34, 63, and 2\% in the Bf-mode.

The Q-mode is characterised by the PC being almost absent. Nonetheless, there is clearly detectable PC emission visible in the single-pulse stacks during some rotations that are otherwise pure Q-mode emission. This might be caused by residual B or Bf-mode emission (i.e.\ the pulsar flickering into the B or Bf-mode for a few rotations) or genuine mode mixing, which we investigate in detail in Sect.~\ref{sec:modemixing}. There is some small but measurable bridge emission between the PC and MP. The MP is symmetric and almost triangular-shaped, except for the characteristic shoulder bump on the trailing edge. The IP is relatively bright (10\% of the total fluence) and strongly asymmetric with two kinks or bumps at the leading edge, which are likely caused by one or two additional emission sub-components, analogue to what we show in our von Mises profile decomposition in Fig.~\ref{fig:pulseprofile}. The total flux density almost doubles ($1.7 \times$) when switching from the Q to the B-mode. However, this is not only because of the appearance of the PC profile component but also because the emission in the MP is strongly enhanced. Specifically, the change is accounted for by a 70\% flux density increase in the MP, a $5 \times$ increase in the PC range, and a halving of the IP's flux density. The PC radiates much more intensely, and a kink at its leading edge, reminiscent of a separate profile sub-component, is visible. The bridge emission is slightly increased. The MP looks almost similar to a scaled-up version of the Q-mode MP, except that the sub-component at the leading edge is enhanced, which results in a steepening of the leading profile edge and a slight shift of the profile maximum to earlier phases. The shoulder hump becomes more apparent. The IP emission is roughly halved, becomes more symmetric, and the central reference point on the peak shifts to slightly later phases. The trailing edge is little affected. The most prominent characteristic of the Bf-mode is that the amplitude or flux density ratio between the PC and MP is closer to being equal at about 1:2. In other words, the PC contributes a significantly higher fraction of flux density than in the Q or B-modes. Indeed, the PC and MP look almost comparable in brightness in the single-pulse stacks. The kink in the PC becomes clearer and most of the additional emission comes from the centre of the PC envelope. The bridge emission near the PC's trailing edge is slightly increased. The brightness of the MP changes little, and so does its shape. The IP is almost unchanged compared with the B-mode.

\subsubsection{Fluctuation analysis}
\label{sec:fluctuationanalysis}

\begin{figure*}
  \centering
  \includegraphics[width=\columnwidth]{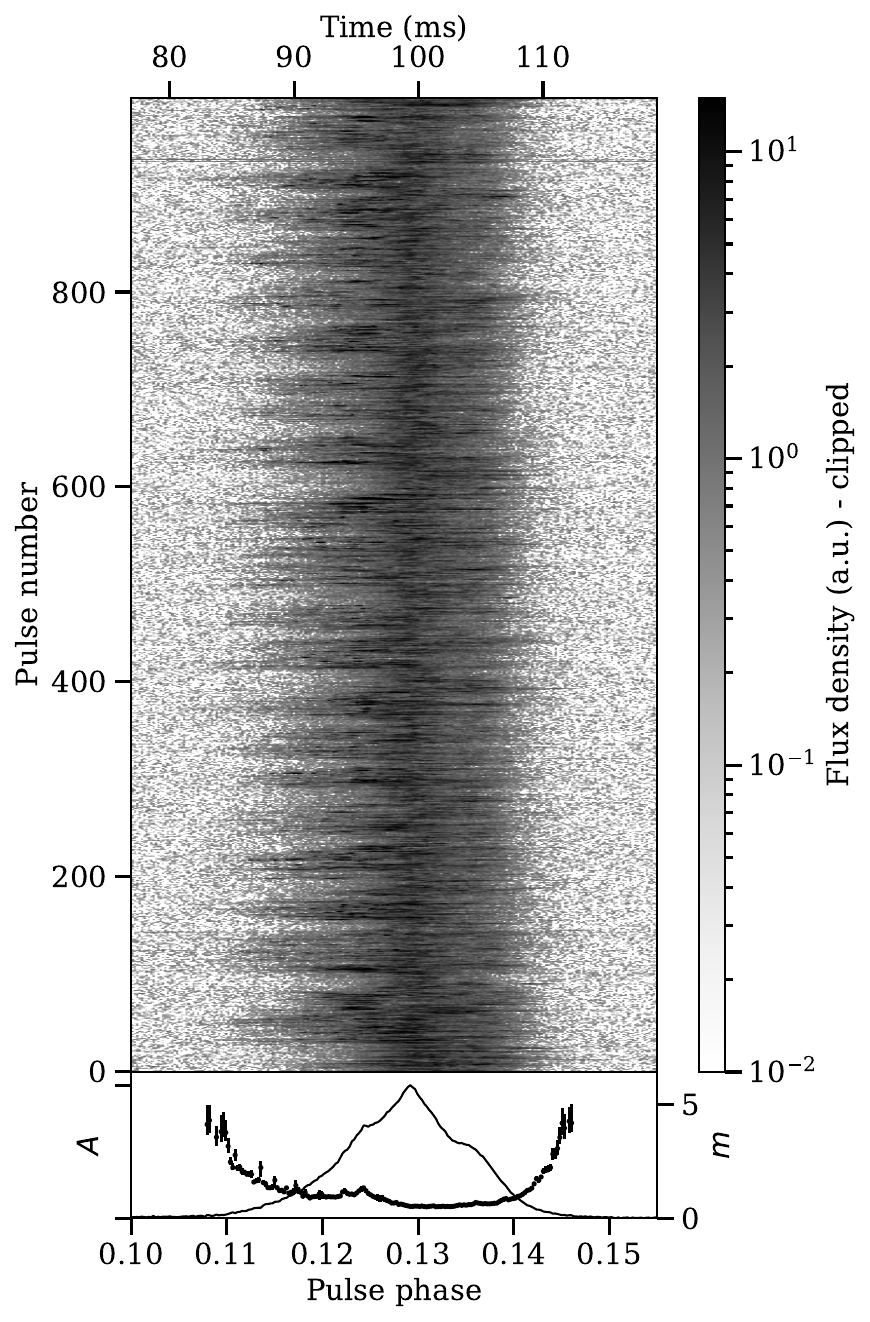}
  \includegraphics[width=\columnwidth]{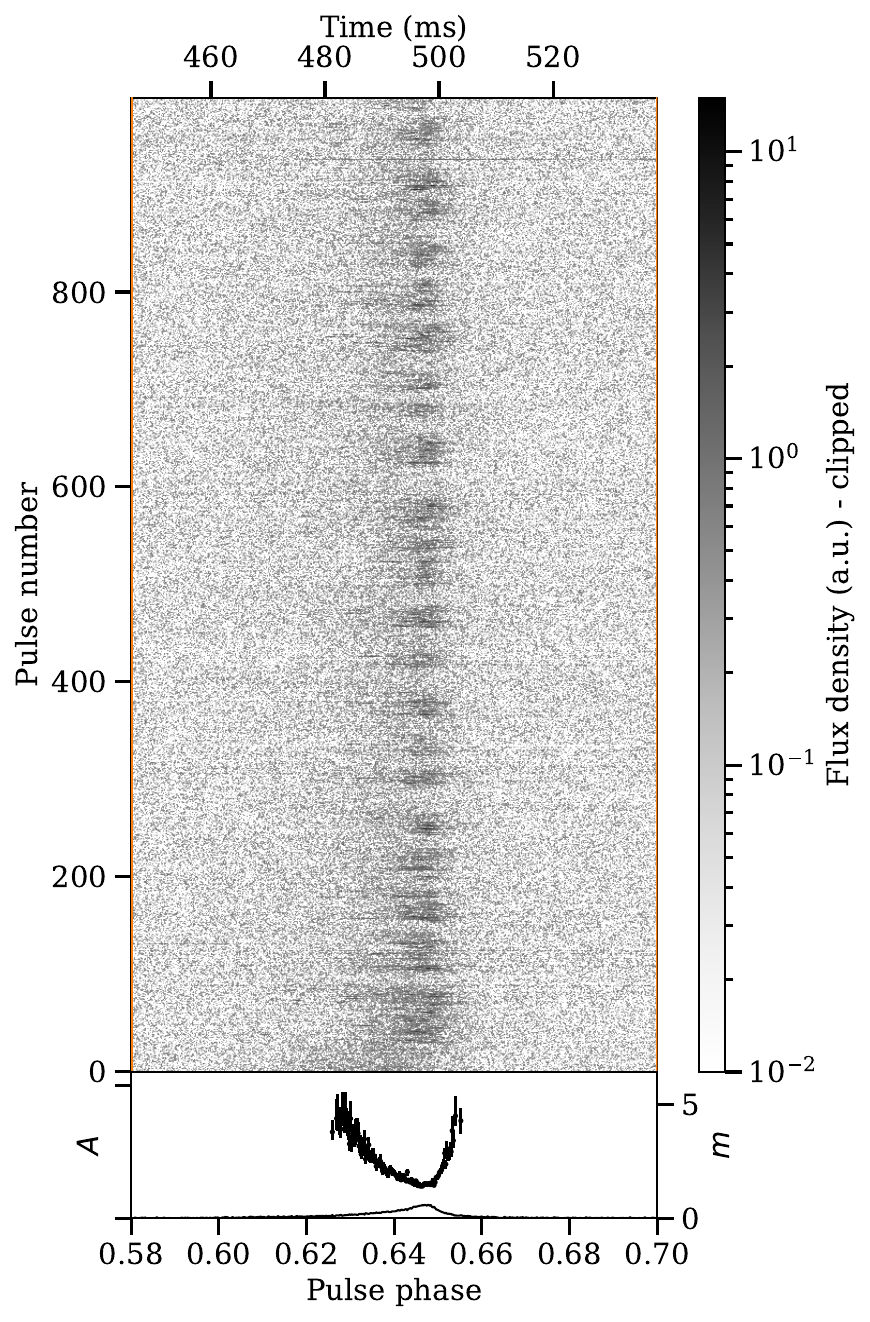}
  \caption{Zoomed-in and dynamic range compressed profile stack of 1000 continuous rotations (pulse numbers 150 to 1150) of Q-mode emission taken from the second observation on 2023-04-24, as shown in Fig.~\ref{fig:singlepulsestacks01} right panel. The bottom panels show the same parameters as in Fig.~\ref{fig:singlepulsestacks01}. We separately display the MP (left) and IP phase range (right). A longitude-stationary amplitude modulation in the leading profile sub-component of the MP is clearly visible. The IP's intensity is modulated with the same period and synchronously with the MP's leading profile component.}
 \label{fig:qmodeampmodulation}
\end{figure*}

\begin{figure}
  \centering
  \includegraphics[width=\columnwidth]{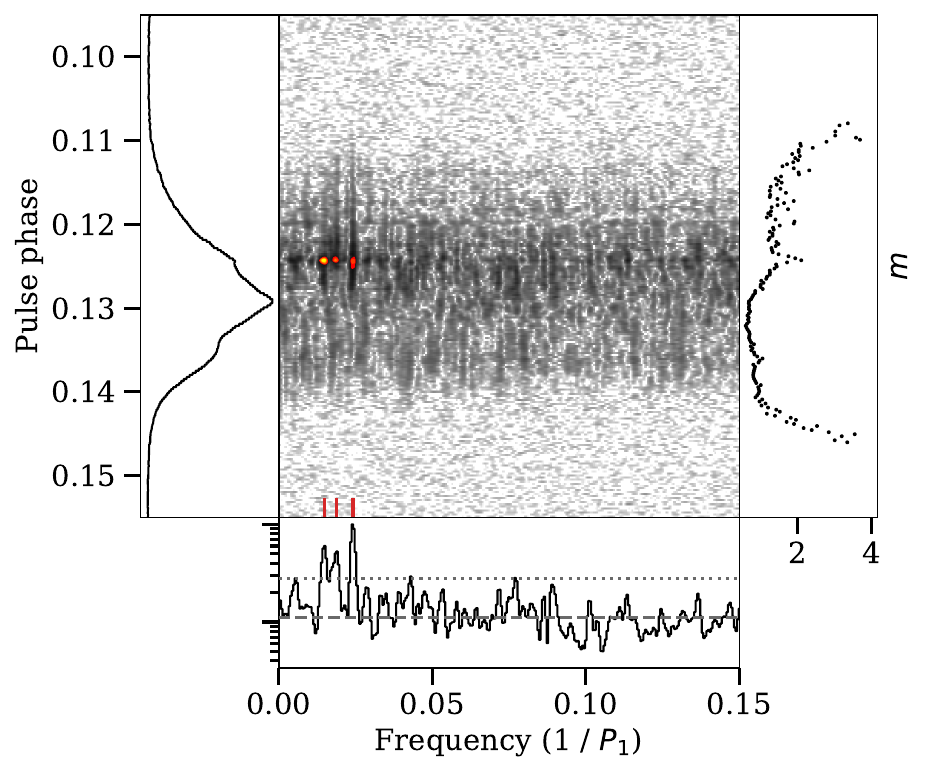}
  \caption{Longitude-resolved fluctuation spectrum of the pulse stack shown in Fig.~\ref{fig:qmodeampmodulation}. The FFT block size was set to 2000 rotations ($2 \times$ oversampling). The panels show the normalised pulse profile (left), the LRFS (middle), the Fourier-domain modulation index (right), and the summed Fourier power on a logarithmic scale (bottom). The red-orange contours are drawn at 0.5, 0.6, 0.7, 0.8, 0.9, and 0.95 of the maximum fluctuation PSD, and we zoomed into the area around the maximum. The lines in the bottom panel show the mean Fourier power (dashed grey line) and the $5~\sigma$ power level (dotted grey line). Three significant $\geq 5~\sigma$ features are visible that align with the leading hump in the MP's profile and the local peak in modulation index $m$. We marked their centroid frequencies with red vertical lines.}
 \label{fig:lrfs}
\end{figure}

\begin{figure}
  \centering
  \includegraphics[width=\columnwidth]{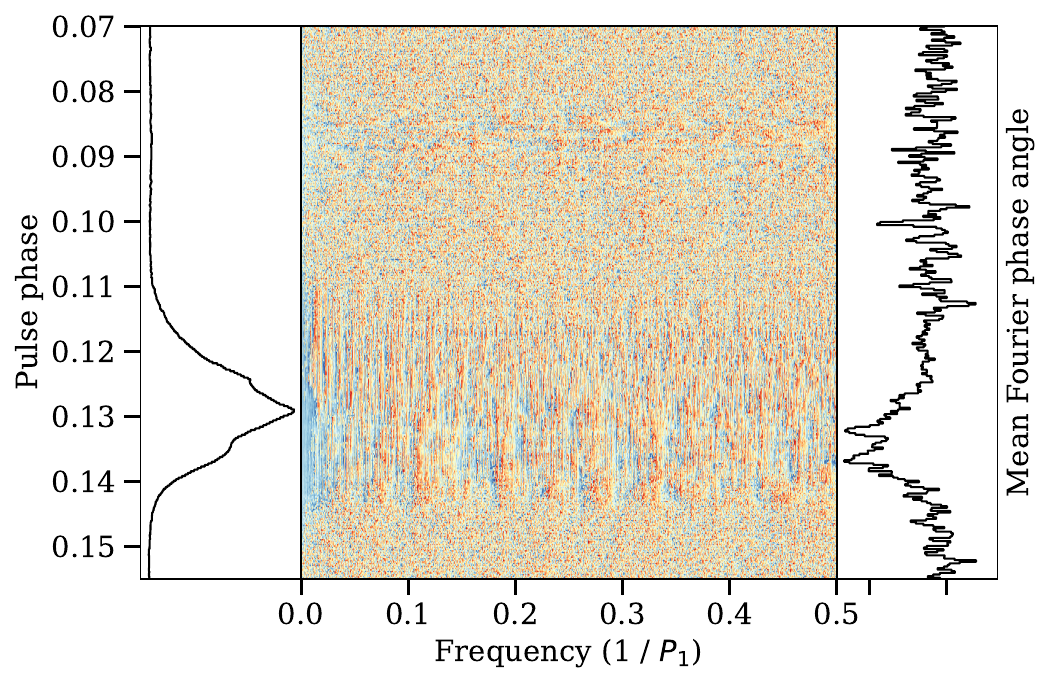}
  \includegraphics[width=\columnwidth]{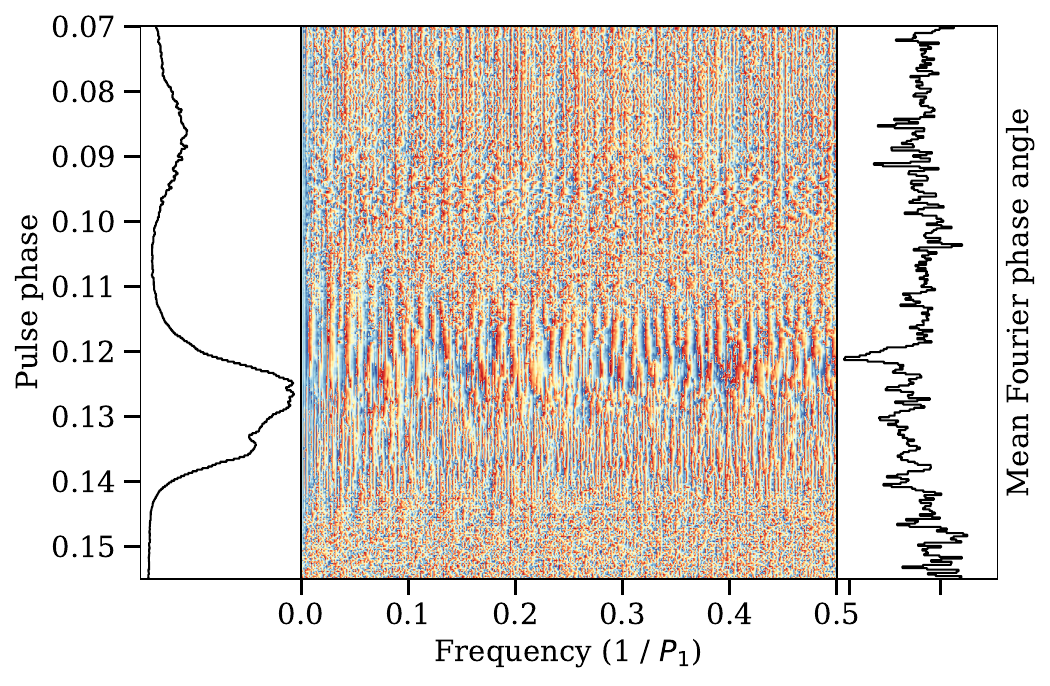}
  \includegraphics[width=\columnwidth]{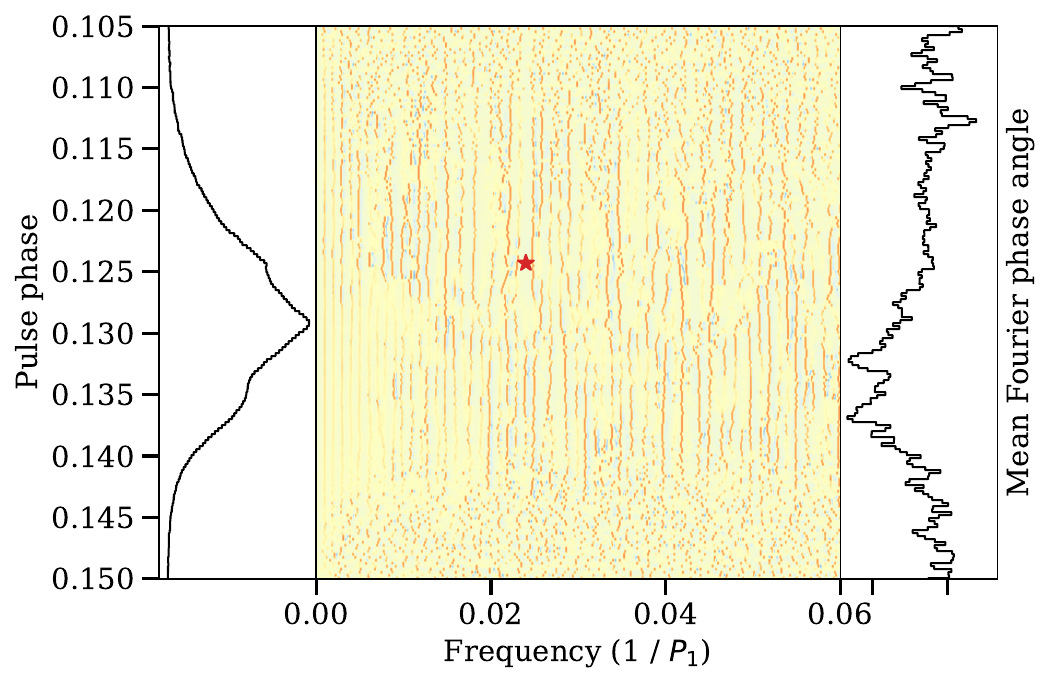}
  \caption{Longitude-resolved phase angle spectra of the Q-mode (top) and the B and Bf-mode (middle). The LRPS visualises the phase relationship between different fluctuation frequencies in the pulsar's emission, in contrast to the LRFS, which shows their magnitudes. The FFT block size was set to 8000 and 1680 rotations, corresponding to eight and four times oversampling. Bottom: phase angle gradient along the frequency axis of the Q-mode LRPS from the top panel. We zoomed into the area near the $P_3 = 41.7 \: P_1$ modulation feature shown with a star marker.}
 \label{fig:lrps}
\end{figure}

We performed a fluctuation analysis of the pulsar's emission by computing a longitude-resolved fluctuation spectrum (LRFS; \citealt{1970BackerLRFS, 1970BackerLRFS2, 1973Backer}) and a two-dimensional fluctuation spectrum (2dFS; \citealt{2002Edwards}) of subsets of the single-pulse stacks. We first identified the longest continuous pulse number ranges during which the pulsar was emitting uninterrupted in one of its modes without switches, based on the single-pulse stacks shown in Figs.~\ref{fig:singlepulsestacks01}--\ref{fig:singlepulsestacks02} and our Markov switching model classification above. For the Q-mode, we selected a stretch of 1000 continuous rotations between pulse numbers 150 and 1150 from the second observation on 2023-04-24, which we show in Fig.~\ref{fig:qmodeampmodulation}. This rotation range is ideal, as it is unusually long and almost RFI-free.

We present the resulting LRFS in Fig.~\ref{fig:lrfs}. The panels show the normalised pulse profile on the left, the LRFS with the median off-pulse spectrum subtracted in the middle, the phase-resolved Fourier-domain modulation index $m$ computed from the LRFS on the right, and the summed Fourier power spectral density (PSD) along the phase bin axis in logarithmic scale at the bottom. The bottom panel with the summed Fourier power also shows the mean Fourier power level across the fluctuation frequencies (dashed grey line) and the $5~\sigma$ power level (dotted grey line) computed using the robust standard deviation. We define LRFS peaks as significant that clearly exceed the $5~\sigma$ power level for more than a single frequency bin. This approach makes it straightforward to identify genuine fluctuation frequencies in the pulsar. We trialled several fast Fourier transform (FFT) block sizes and window functions. A Hann window and FFT block length of 2000 rotations, that is, $2 \times$ oversampling the data by zero padding, provides the most suitable frequency resolution.

Three significant $\geq 5~\sigma$ feature clumps are visible in the MP phase region. The left-most feature in Fig.~\ref{fig:lrfs} (feature~1) contains the maximum fluctuation power per bin and corresponds to a fluctuation period $P_3 = (66.7 \pm 2) \: P_1$. It aligns perfectly with the leading bump or shoulder in the MP profile. The feature is slightly extended in both pulse phase (15 phase bins) and towards lower fluctuation frequencies (7 FFT bins). The middle feature (feature~2) at $P_3 = (52.6 \pm 1) \: P_1$ is narrow in fluctuation frequency (4 FFT bins) and fairly extended in pulse phase (55 bins). It seems to merge with feature~1 at lower frequencies. The right-most feature (feature~3) at $P_3 = (41.7 \pm 0.9) \: P_1$ contains the maximum integrated power. It is reasonably narrow and well-defined in fluctuation frequency (4 bins) but extends significantly in pulse phase ($\sim$73 bins). Significant fluctuation power is visible between the pulse phases 0.11 and 0.128, which correspond to the onset of the MP profile at the leading edge and its profile peak. Interestingly, this FFT bin also contains excess power around the peak of the IP profile. Outside those phase ranges, the power in this FFT bin is consistent with the off-pulse noise. In fact, all three modulation features are replicated at lower intensity in the IP's profile peak. The maximum power bins of all three features align in phase with each other, the leading hump in the MP's profile, and the local maximum in modulation index. Hence, they seem to be genuinely different fluctuation periods in the same profile sub-component. There is no significant fluctuation power in the off-pulse phase range, which indicates that instrumental effects are well suppressed. We quote all the fluctuation periods above at the centres of the maximum amplitude FFT bins and state uncertainties corresponding to one FFT bin width.

Our 2dFS analysis nicely recovers the $P_3 = 41.7 \: P_1$ feature and its harmonics in the MP. The feature is narrow in $P_3 / P_1$ space (2 bins) and slightly extended in $P_2$ space (5 bins). Its fluctuation power peaks in the first non-DC bin in the $P_2$ domain; that is, its maximum is offset by only one bin from the vertical axis. In other words, its $P_2$ is consistent with zero. Therefore, the feature is most likely caused by longitude-stationary amplitude modulation, and not a sub-pulse drifting (phase) effect. The amplitude modulation or oscillation of pulse intensity in the MP's leading profile sub-component is clearly visible in the zoomed-in pulse stack shown in Fig.~\ref{fig:qmodeampmodulation} left panel. We separately performed a 2dFS of the IP, and the $P_3 = 41.7 \: P_1$ modulation is also present there. Both the MP's leading component and the IP's overall intensity are modulated with the same period and synchronously with each other; that is, positively correlated in flux density or fluence (Fig.~\ref{fig:qmodeampmodulation}). This phase-locked relationship between the MP and IP within the Q-mode is an interesting phenomenon. It has been debated whether the MP and IP modulation patterns are phase offset in rotation \citep{2010Backus, 2012Latham, 2019Yan}. To test this, we computed the cross-correlation of the MP and IP fluence time series extracted from the data shown in Fig.~\ref{fig:qmodeampmodulation}, with the phase gates set to 0.1--0.155 (MP) and 0.63--0.66 (IP). The cross-correlation peaks at zero lag with a value near 0.4. The $P_3 = 41.7 \: P_1$ periodicity and its harmonics are visible in the cross-correlation and the autocorrelations of the individual time series. This confirms that the fluences of the leading MP profile sub-component and the IP are modulated synchronously with each other and without phase offset, that is, perfectly in phase. In other words, if the fluence of the MP is high, the fluence of the IP directly following it in the same rotation will also be high in a relative sense (positive correlation).

We repeated our analysis for the pulsar's B and Bf-mode emission. We selected the pulse number range from 2990 to 3410 of the second observation on 2024-04-24, as shown in Fig.~\ref{fig:singlepulsestacks01} right panel. This is the longest continuous B and Bf-mode emission stretch in our data. We averaged up to eight oversampled LRFS periodograms using Welch's algorithm \citep{1967Welch} with a Hann window to reduce and smooth the noise floor. There are several peaks visible in the LRFS below $0.11 \: P_1^{-1}$ whose maximum power bins align with a subtle bump (phase range 0.121--0.124) at the leading edge of the MP's profile just before the first local peak. However, they are only marginally above the integrated noise floor ($\leq 3~\sigma$). Aside from those, there is no clear periodicity visible in the LRFS. The 2dFS confirmed the absence of any obvious modulation.

To investigate the single-pulse modulation further, we computed what we call a `longitude-resolved phase angle spectrum' (LRPS), which is analogue to the LRFS. It visualises the phase angle of the complex Fourier coefficients for all fluctuation frequencies and phase bins. Hence, it contains complementary information to an LRFS analysis, which only considers the magnitudes of the complex Fourier components. The mean complex off-pulse spectrum was subtracted before computing the angles. We show the resulting LRPS in Fig.~\ref{fig:lrps} for the Q-mode and the B and Bf-mode pulse stacks described above. We oversampled the data by zero-padding the FFT by a factor of eight and four times, respectively. We wanted to achieve a high fluctuation frequency resolution to investigate the three features identified in the LRFS and to characterise and perhaps distinguish them in the phase angle domain. The higher frequency resolution helps the visual inspection. However, it also resolves the response of the Fourier bins. The panels in Fig.~\ref{fig:lrps} show the normalised pulse profile, the LRPS, and the mean Fourier phase angle computed along the fluctuation frequency axis. We mapped negative phase angles to blue shades, zero to yellow, and positive angles to red shades. The red and orange horizontal lines denote the on-pulse phase ranges as in the other figures. The LRPS shows the phase relation between different Fourier components, that is, how they align in phase with respect to each other.

The MP LRPS are strikingly different between the modes. The Q-mode (Fig.~\ref{fig:lrps} top panel) shows relatively large patches of similar phase angles adjacent in bin and frequency space. The $P_3 = 41.7 \: P_1$ modulation feature at $0.024 \: P_1^{-1}$ sits right in the middle of one of the `ridge lines' at approximately zero phase angle in a low negative gradient ($-25$~deg) area (star marker in Fig.~\ref{fig:lrps} bottom panel). There is a large patch of mostly negative phase angle at fluctuation frequencies below $0.012 \: P_1^{-1}$ that extends from the MP's profile leading hump to its trailing flank. A similar patch is also visible in the IP. This low-frequency red noise is most likely due to scintillation or a pulsar intrinsic coherent noise process. Interestingly, that area appears to be smoothed out to an almost constant angle. The mean phase angle is noticeably negative across the MP, particularly starting from the leading hump until the trailing end. The maximum negative excursions of $-13$~deg happen symmetrically around the trailing saddle point in the profile. There is also a hint of low-level PC emission visible at small fluctuation frequencies $\leq 0.1 \: P_1$. In contrast, the B and Bf-mode LRPS (Fig.~\ref{fig:lrps} middle panel) shows rapid phase angle variation with fluctuation frequency. The MP's trailing edge looks unstructured and almost randomised, while the leading edge exhibits narrow, regularly occurring patches of rapid negative phase angle sweep. No clear structures are visible in the PC apart from the negative angle smoothing at low fluctuation frequencies. The angles change rapidly with frequency. The IP LRPS looks qualitatively similar between the modes. The difference is primarily in patch intensity and that the Q-mode looks smoother. Overall, the phases change more smoothly in the Q-mode, whereas the B and Bf-mode shows more rapid phase angle variation and appears more chaotic. This agrees with our intuition gained from looking at the pulse stacks and single-pulse profiles.

\subsubsection{Correlated mode switching of the PC, MP, and IP}
\label{sec:correlation}

\begin{figure}
  \centering
  \includegraphics[width=\columnwidth]{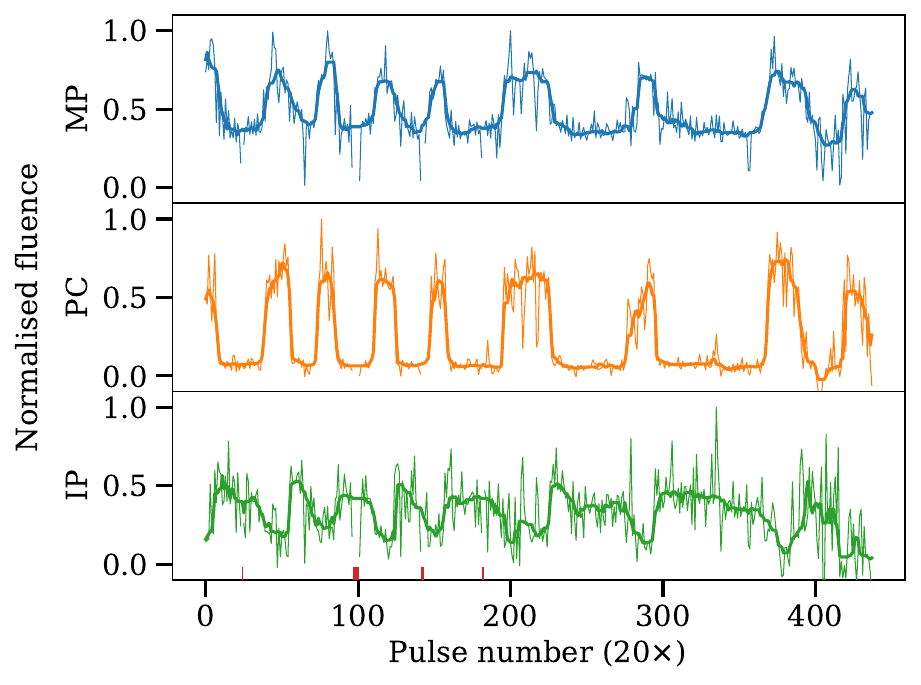}
  \caption{Correlation analysis between the emission in the MP, PC, and IP phase regions. We show the time series of the measured on-pulse fluences normalised by their individual maxima, where we averaged every 20 pulses for clarity. The solid lines depict running median smoothings of the data using ten-sample wide sliding windows. The red markers at the bottom highlight integrations that were excised. The MP and PC components switch fluence synchronously to each other, while the IP switches asynchronously to both the PC and MP. The latter strongly suggests a feedback mechanism between the polar caps on opposite sides of the star.}
 \label{fig:componentcorrelation}
\end{figure}

When looking at the single-pulse stack in a logarithmic greyscale (Figs.~\ref{fig:singlepulsestacks01}--\ref{fig:singlepulsestacks02}), we noticed that the IP emission is wider and somewhat brighter when the pulsar is in the Q-mode in the MP region. When the B or Bf-mode mode is active in the MP, the IP is significantly narrower and fainter. We investigated this by extracting time series of fluence and boxcar equivalent pulse width separately in the PC (0.01--0.107), MP (0.107--0.2) and IP (0.58--0.7) phase ranges. These are shown in Fig.~\ref{fig:componentcorrelation}, where we averaged every 20 pulses to increase the S/N. To reduce the noise further, we computed median smoothings of the time series with 10-sample (200 rotation) wide sliding windows. Those are shown as solid lines. The MP shows an almost constant fluence in the Q-mode, which approximately doubles when it enters the B or Bf-mode. This is not just because of the PC component's appearance; the MP actually gets brighter in both peak and component averaged flux density. At the same time, the PC fluence increases by a factor of $\sim$10 from being almost consistent with the baseline noise. The dynamic range is significantly higher than in the MP. The IP fluence timeline is much noisier because of its small absolute value but increases by a factor of $\sim$3 after the pulsar switches to the Q-mode. Figure~\ref{fig:componentcorrelation} also shows the almost perfect correlation between the moding in the different profile components. The PC and MP switch fluence synchronously to each other and asynchronously to the IP. The same behaviour can be seen in the boxcar equivalent pulse width. This means there is a feedback between the emission emanating from opposite poles of the star. The switching of the magnetosphere between the stable emission configurations on one side of the star must interact with that at the opposite magnetic pole. Hence, a feedback mechanism and information transfer between both polar caps must exist, as was pointed out early on \citep{1982Fowler, 2005DyksA}.

%
%
\subsection{Single-pulse profile morphology, dynamic spectra, and microstructure}
\label{sec:spmorphology}

\begin{figure*}
  \centering
  \includegraphics[width=0.33\textwidth]{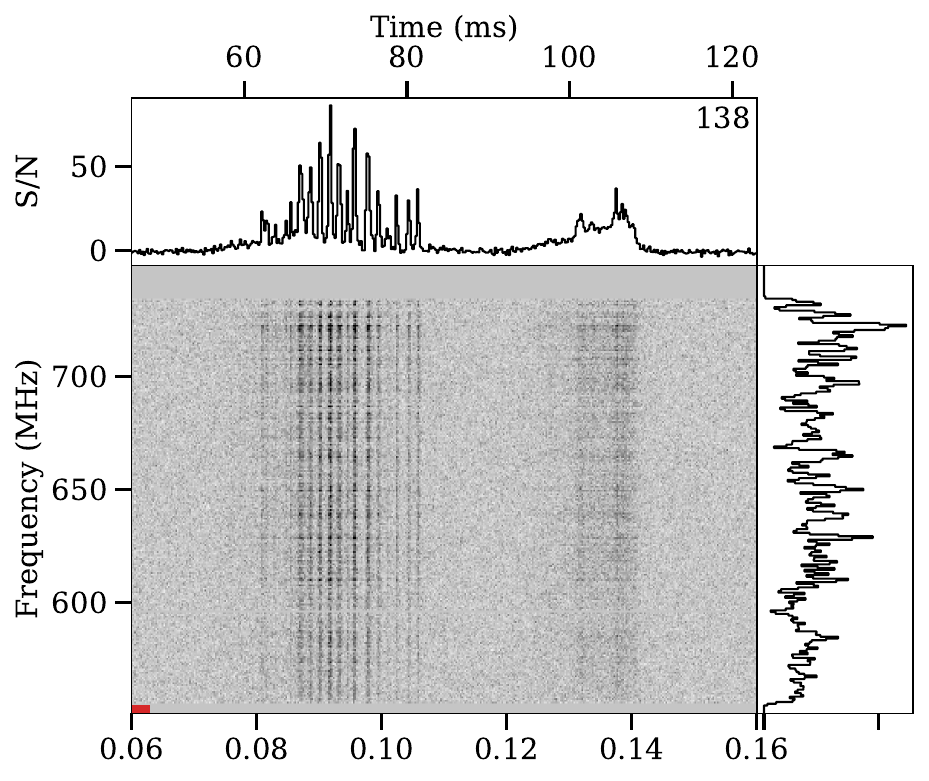}
  \includegraphics[width=0.33\textwidth]{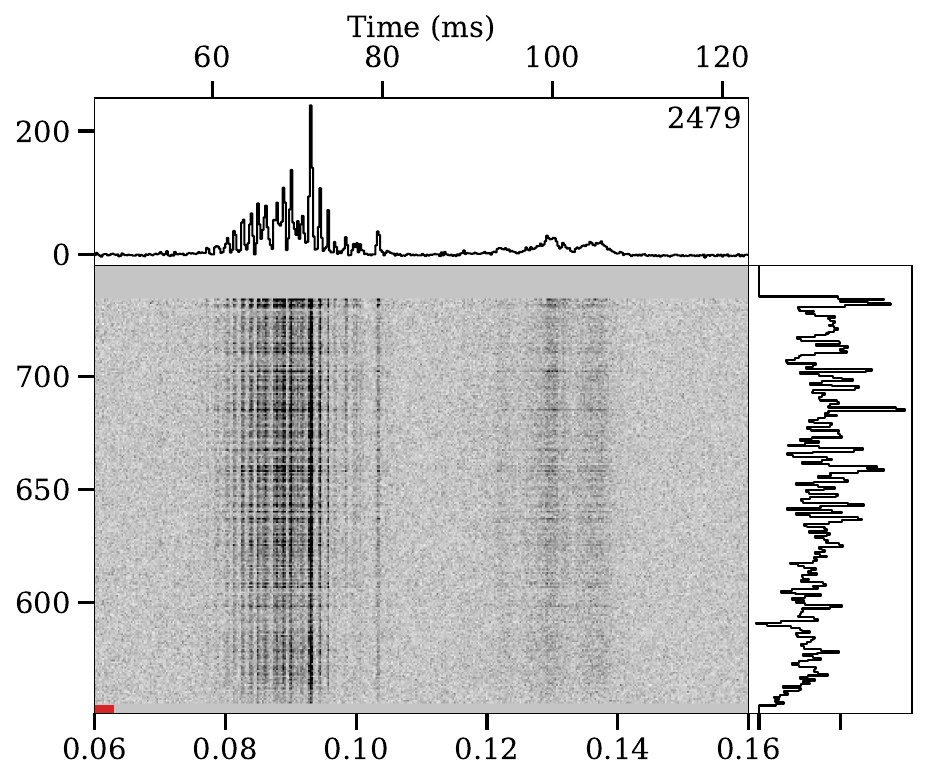}
  \includegraphics[width=0.33\textwidth]{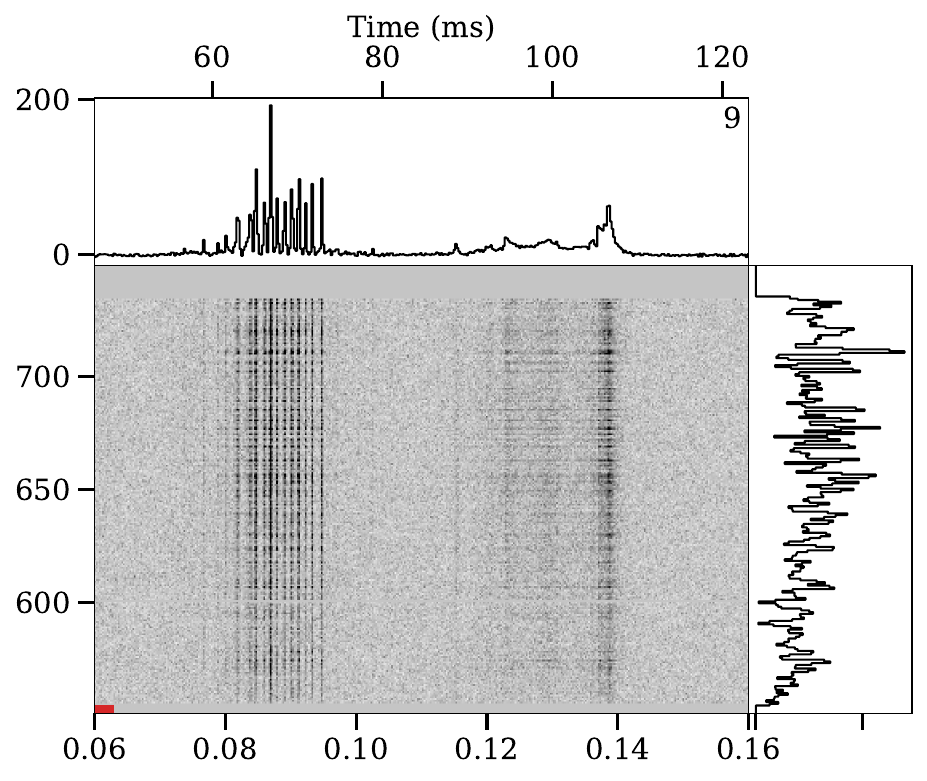}
  \includegraphics[width=0.33\textwidth]{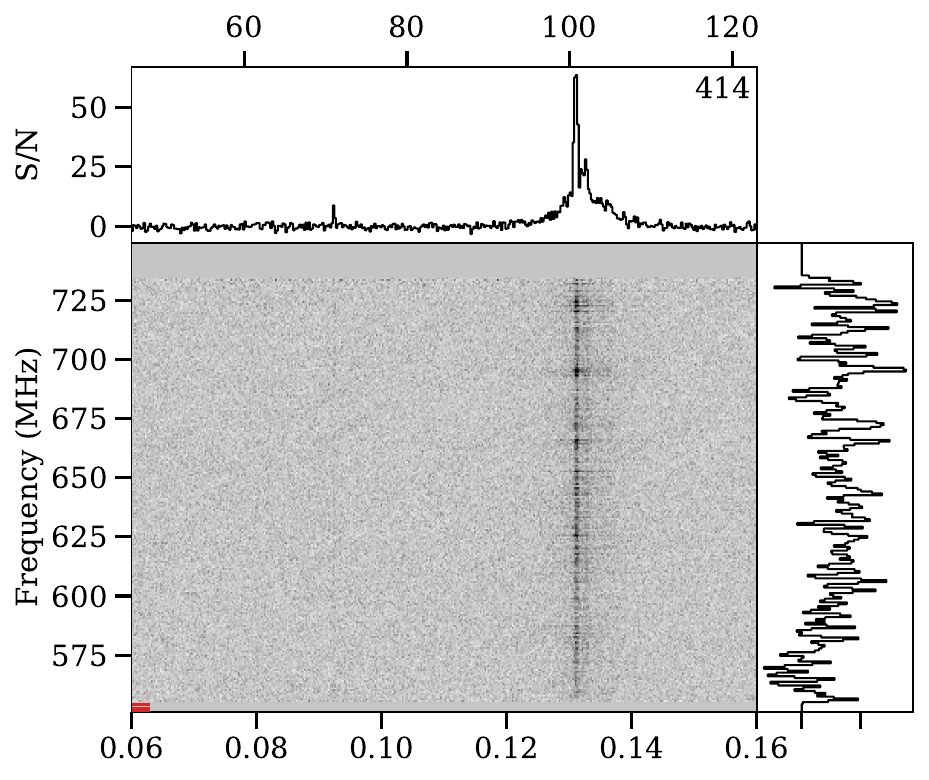}
  \includegraphics[width=0.33\textwidth]{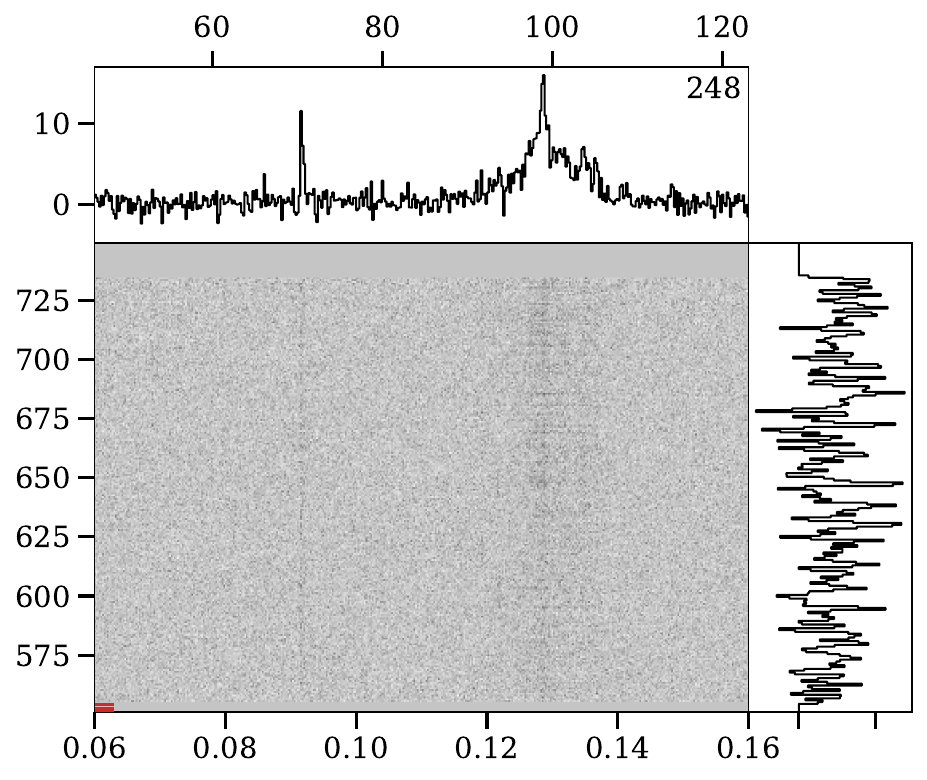}
  \includegraphics[width=0.33\textwidth]{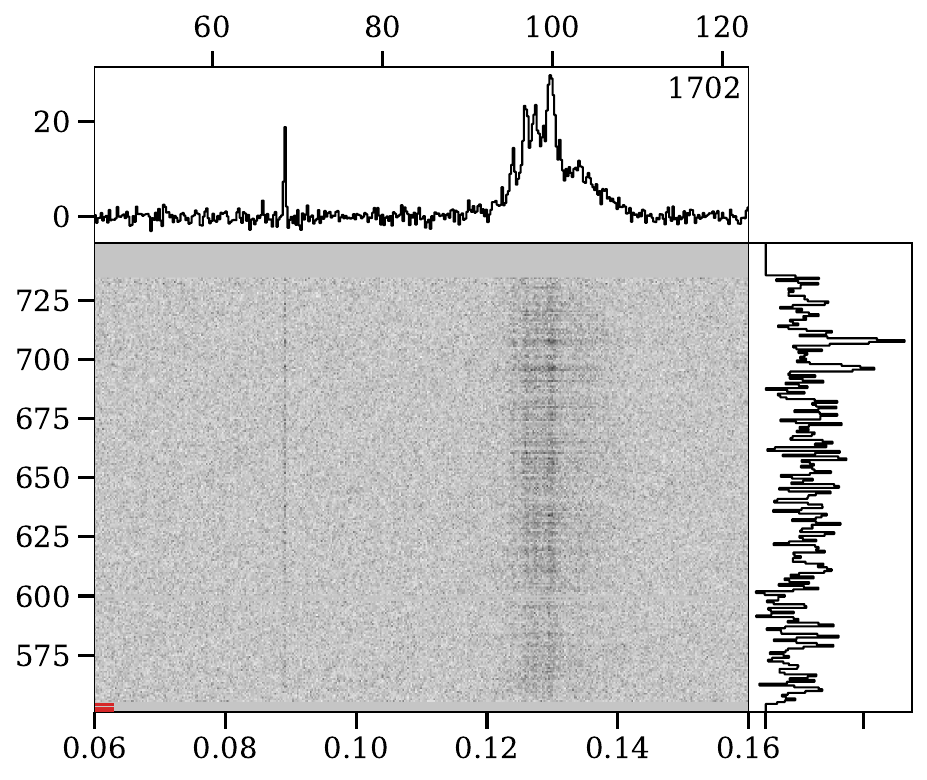}
  \includegraphics[width=0.33\textwidth]{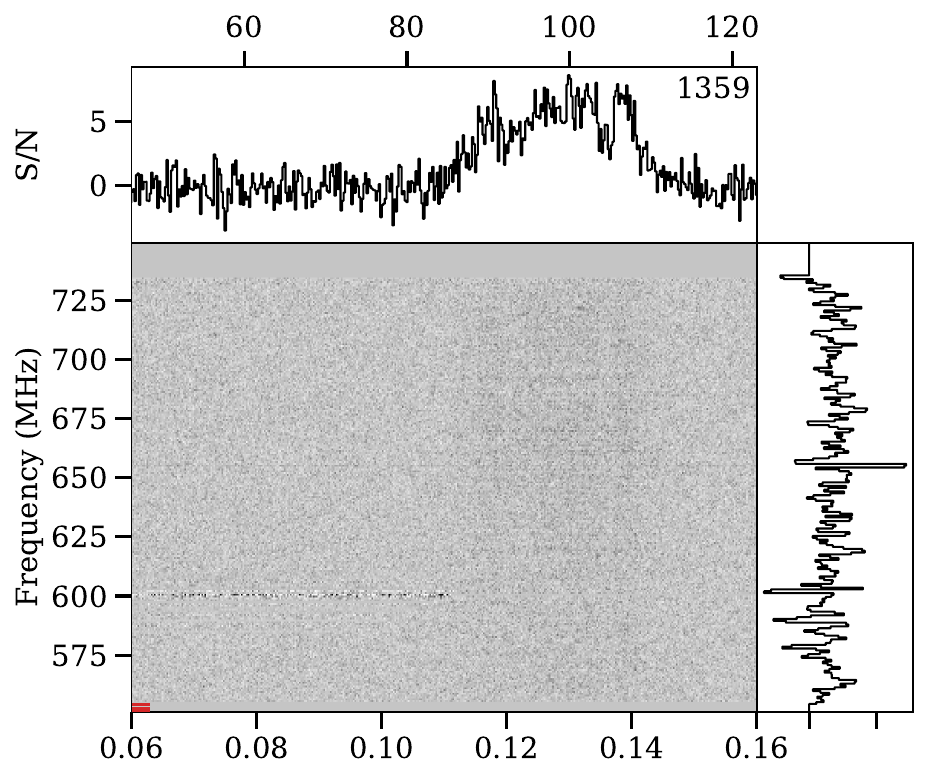}
  \includegraphics[width=0.33\textwidth]{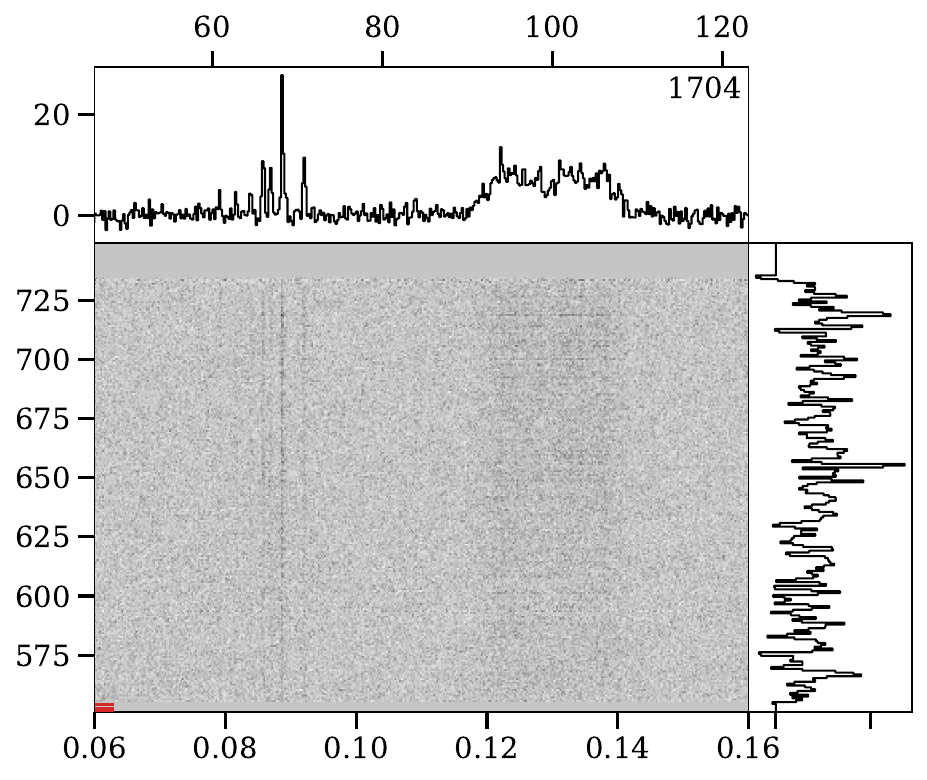}
  \includegraphics[width=0.33\textwidth]{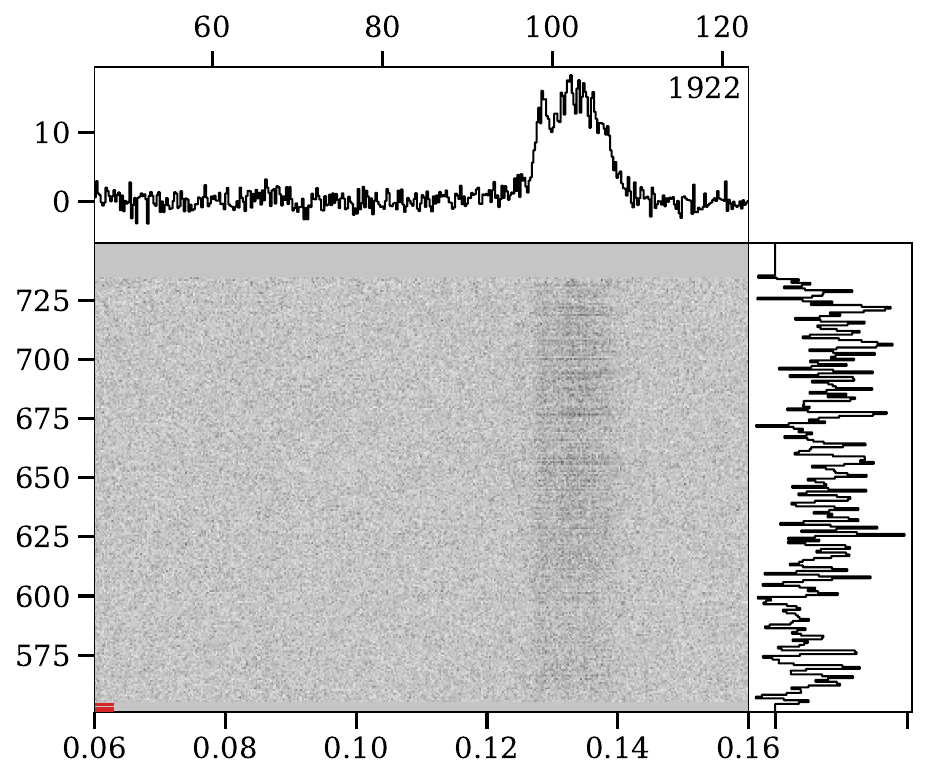}
  \includegraphics[width=0.33\textwidth]{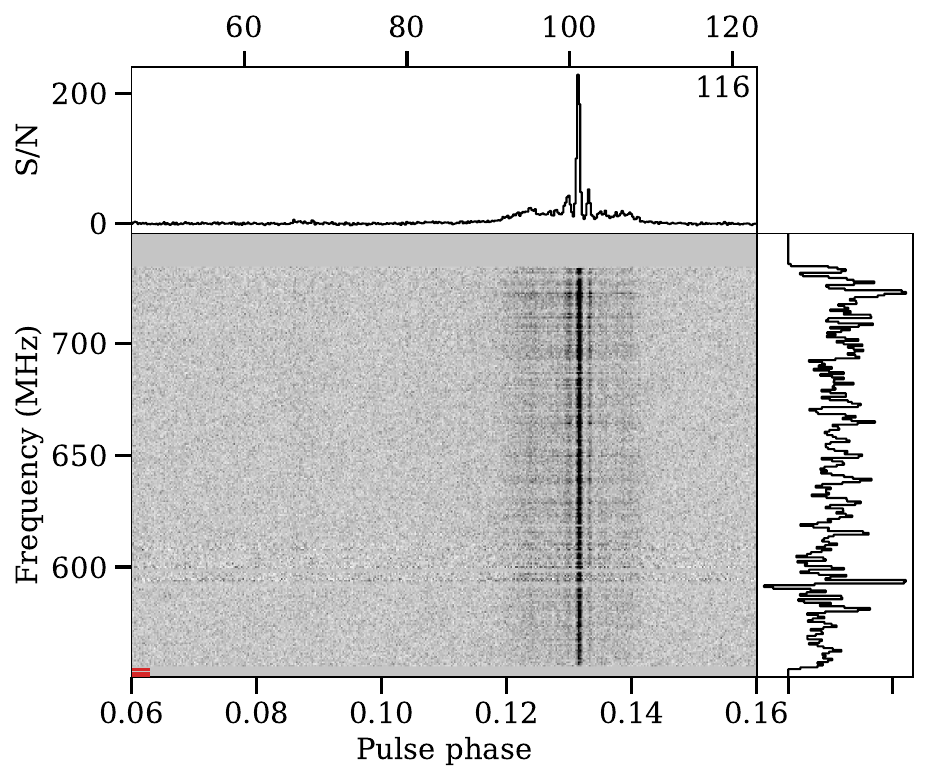}
  \includegraphics[width=0.33\textwidth]{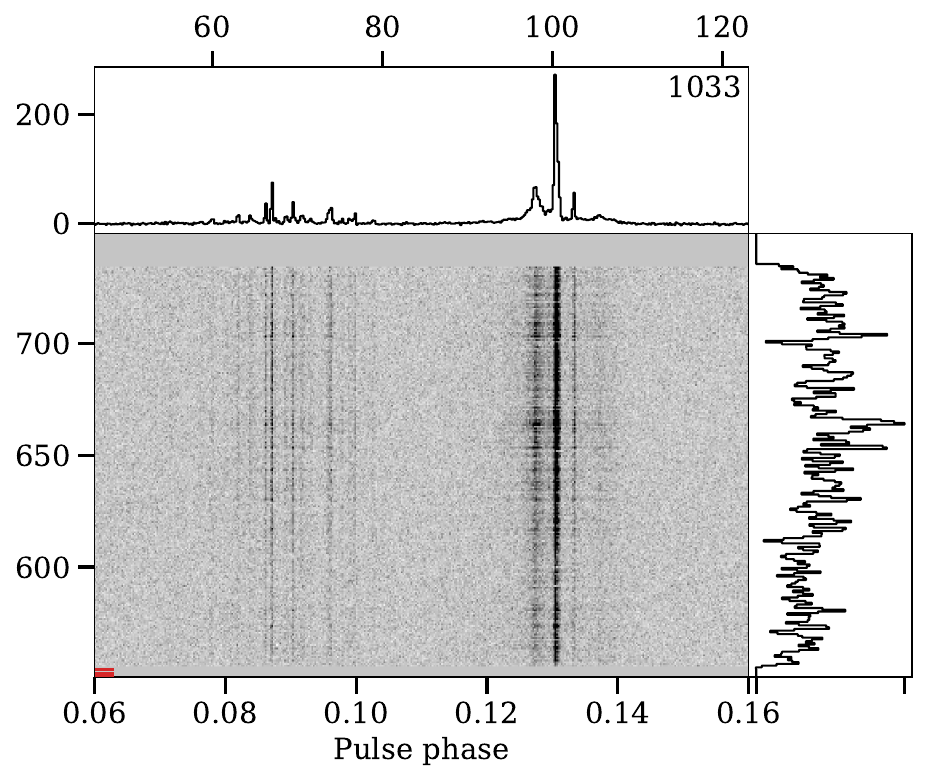}
  \includegraphics[width=0.33\textwidth]{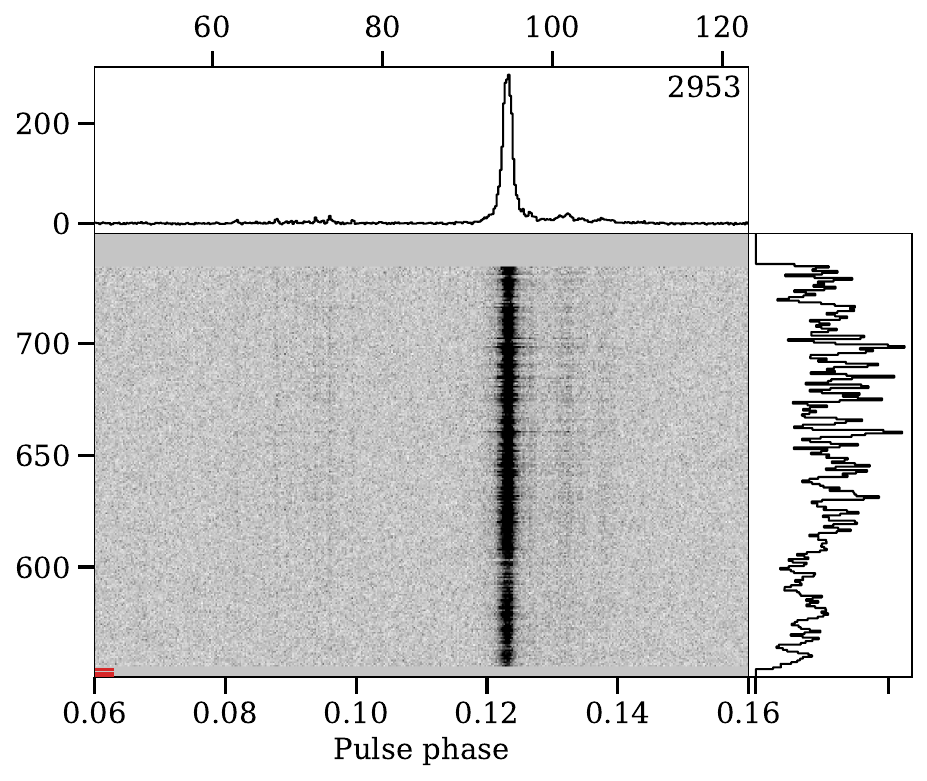}
  \caption{Three representative examples of single pulses exhibiting extensive quasi-periodic microstructure (top row), low-level PC activity within long stretches of Q-mode emission (second row), low-intensity square-like pulses (third row), and extremely bright pulse components (bottom row) as observed in the PC and MP phase range. We display the pulse numbers in the top-right corner of the profile panels. The dedispersed dynamic spectra are saturated at the same intensity in each panel.}
 \label{fig:pulsemorphology}
\end{figure*}

To characterise PSR~B1822$-$09's single-pulse profile morphology and emission characteristics, we visually inspected the 8743 single pulses in the observation taken on 2023-04-24. We found several peculiar single-pulse phenomena in the data that we describe below. In Fig.~\ref{fig:pulsemorphology}, we show three representative examples of single pulses of each phenomenon.

\subsubsection{Quasi-periodic microstructure in the PC component}
\label{sec:microstructure}

\begin{figure*}
  \centering
  \includegraphics[width=\columnwidth]{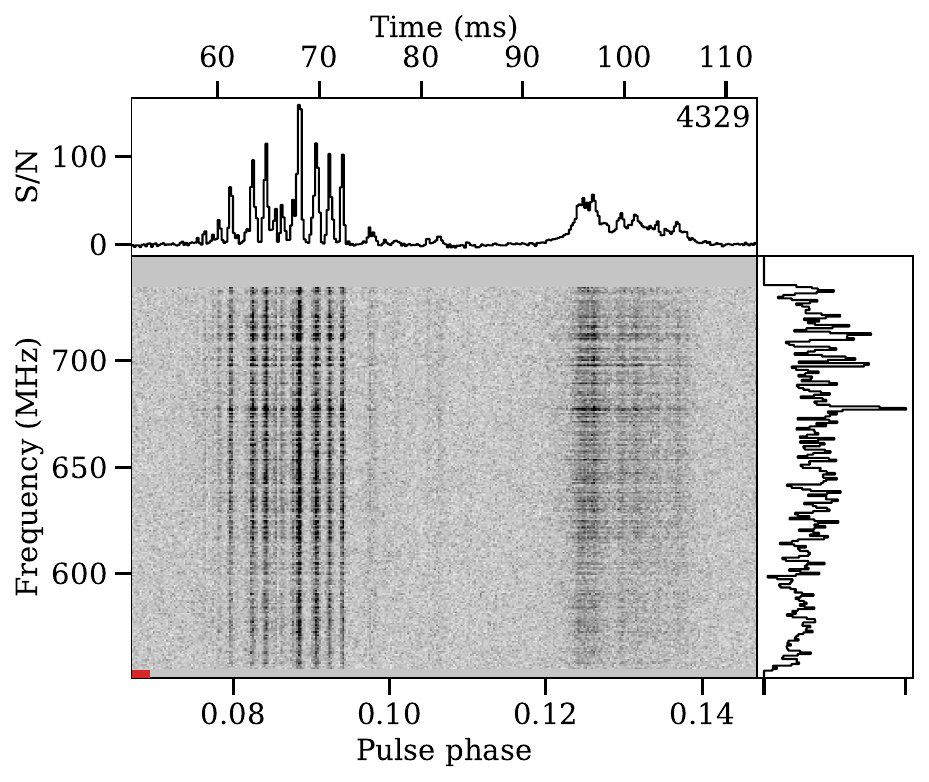}
  \includegraphics[width=\columnwidth]{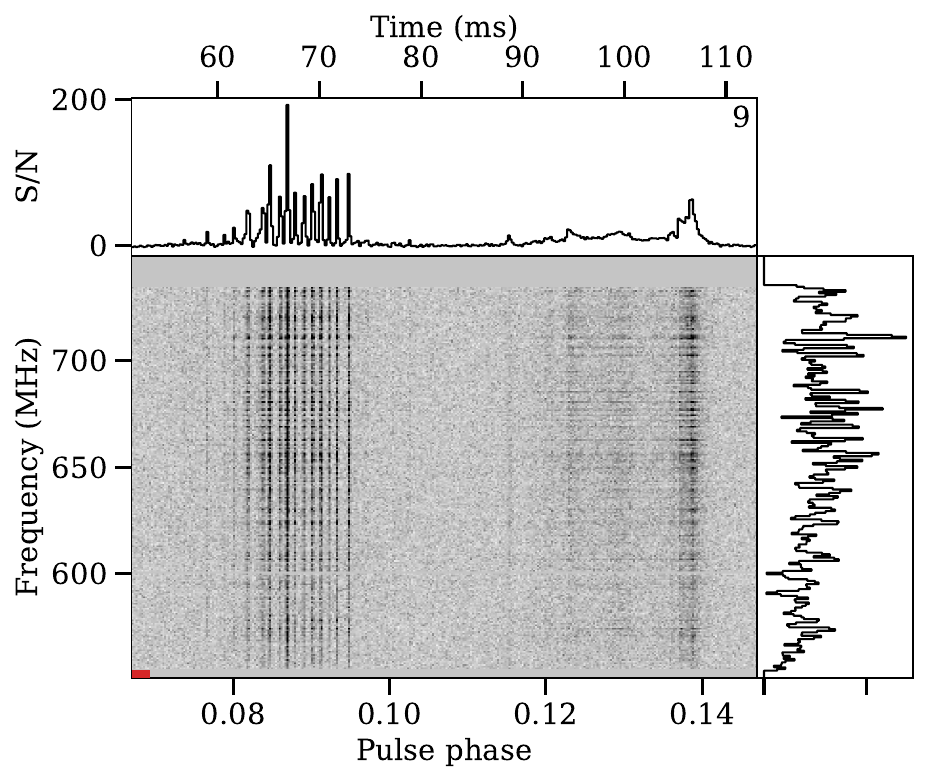}
  \caption{Dedispersed dynamic spectra of two example single pulses that exhibit extensive quasi-periodic microstructure in the PC profile component. We zoomed into the PC and MP phase range as in Fig.~\ref{fig:pulsemorphology}. The quasi-periodic microstructure is clearly visible in the PC pulse envelope.}
 \label{fig:pcmicrostructure}
\end{figure*}

\begin{figure}
  \centering
  \includegraphics[width=\columnwidth]{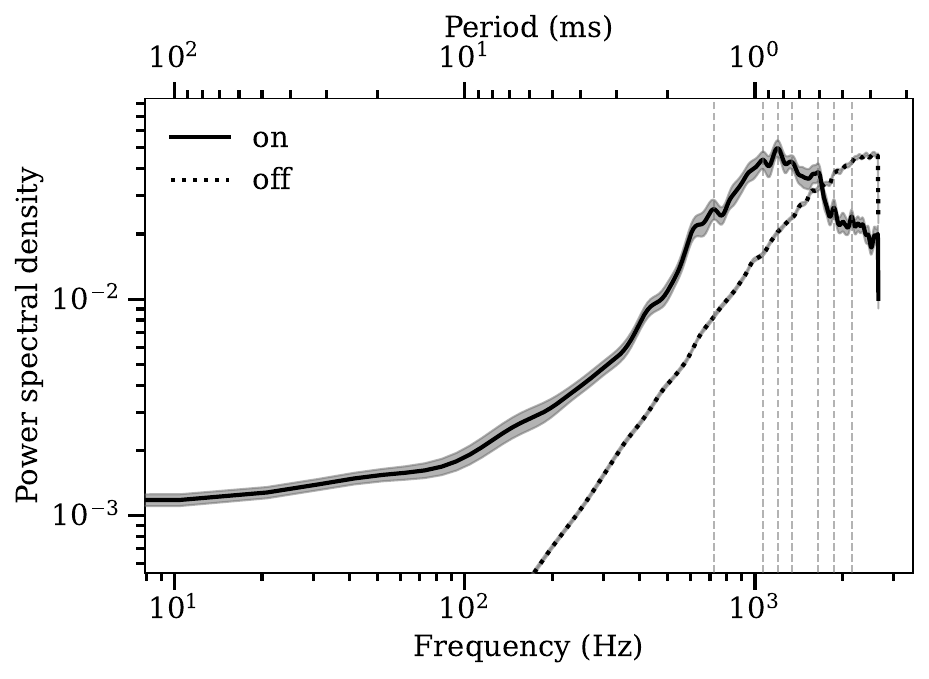}
  \caption{Fourier power spectral density averaged over 59 single pulses that exhibit extensive quasi-periodic microstructure in the PC profile component. The black solid line shows the sample mean, and the grey shaded area its standard error for the combined PC and MP on-pulse phase range. The black dotted line is the same for the off-pulse region. The grey dashed lines mark significant features. The dominant feature occurs near 1200~Hz or 0.83~ms.}
 \label{fig:quasiperiodpsd}
\end{figure}

The most striking feature is the quasi-periodic microstructure seen primarily in the PC pulse profile component when the pulsar is in the Bf-mode (Fig.~\ref{fig:pulsemorphology} top row). We show additional example single pulses that exhibit extensive PC microstructure in Fig.~\ref{fig:pcmicrostructure}. As can be seen, the PC consists of extremely narrow (1--2 bins wide) shot-like micropulse emission that can be very bright with up to S/N $\geq 200$ per phase bin. The micropulses often occur quasi-periodically within the PC phase window, as is evident from Figs.~\ref{fig:pulsemorphology}--\ref{fig:pcmicrostructure}. We analysed their quasi-periodicity in the frequency domain by computing the Fourier power spectral density (PSD) using Welch's algorithm with a Hann window on the first difference of the pulse profile time series, that is, $\delta_i = A_{i+1} - A_i, \forall i \in [0, N_\text{on} - 1]$, where $A_i$ is the Stokes I profile amplitude at the phase bin $i$, and $N_\text{on}$ is the number of on-pulse phase bins. Taking the profile difference is beneficial, as it suppresses any low-frequency red noise. The differencing operation effectively acts as a high-pass filter. Additionally, we performed an autocorrelation function analysis for comparison and computed time-resolved spectrograms using the short-time Fourier transform and wavelet scalograms \citep{1998Torrence} to understand the temporal evolution in frequency across the pulse profile. We then stacked and averaged the Fourier PSDs of 59 single-pulses that show strong PC microstructure. Figure~\ref{fig:quasiperiodpsd} shows the resulting mean PSD and its standard error for the combined PC and MP on-pulse phase region (on) and the off-pulse phase range (off) for comparison. We used the combined PC and MP on-pulse region to have a sufficiently large number of phase bins available for the analysis. We highlighted several visually identified PSD features (local maxima) with vertical grey dashed lines. These occur near 720, 1064, 1197, 1337, 1652, 1872, and 2154~Hz, which corresponds to periods of 1.39, 0.94, 0.84, 0.75, 0.61, 0.53, and 0.46~ms. The quasi-periodicity near 1197~Hz (0.84~ms) clearly dominates the PSD. This means that the micropulses typically occur roughly every 0.8~ms within the PC envelope, which agrees well with our visual profile inspection. The mean off-pulse PSD is smooth and featureless from 8~Hz up to the Nyquist frequency, which indicates that instrumental effects are well suppressed. Thus, our Fourier analysis unambiguously confirmed the quasi-periodic emission.

We were then interested in the temporal extent and separation of the micropulses. We investigated this by automatically identifying the peaked profile components above 10 S/N in the time domain and measuring their number, S/N, half-prominence widths, and separations for the same set of single-pulses with extensive PC microstructure as before and as visualised in Fig.~\ref{fig:profilefeatures}. The number of micropulses in the PC and MP phase window ranges between 14 and 34, with a peak in histogram counts between 19 and 21. The full-width half-prominence histogram has a broad top with almost equal counts between 0.22 and 0.41~ms, corresponding to 1.2 and 2.2 phase bins, and falls off rapidly afterwards. The mode of the distribution of pulse separations (centre-to-centre) between consecutive profile peaks lies between 0.73 and 1.08~ms, or 3.9 and 5.8 phase bins, and falls off quickly. However, there is a minor bump near 2.9~ms (15.5 bins), which is probably due to the microstructure within the MP. Additionally, there is a very slight and broad accumulation of counts between 16 and 27~ms, which are the separations between the last peak of the PC and the first of the MP, that is, the PC-MP profile component separation. The S/N histogram shows a maximum S/N per phase bin of 240 with an almost exponential increase towards lower values and a mode near 12. Interestingly, the phase resolution of our single-pulse folded data (187.74~$\mu \text{s}$) seems to resolve the majority of the micropulses. The mode of the width distribution is appreciably offset from the resolution limit, and the normalised bin count approximately halves before reaching the threshold. However, the broad, flat top of the distribution might point towards a pile-up. The native time resolution of our data (81.92~$\mu \text{s}$) will allow for a more detailed investigation in the future. Our analyses in the frequency and time domain agree nicely.

In a subsequent step, we examined where in the single-pulse stacks the pulses with strong PC microstructure happened and whether they were temporally related to any of the other single-pulse phenomena described below. They appear only during the B or Bf-mode and typically occur in small clusters or bursts where two to four pulses show extreme microstructure in almost consecutive pulses. In fact, there are several occasions where we observed strong PC microstructure in truly consecutive pulses. Some micropulses remain phase aligned in those consecutive pulses, that is, static in phase between rotations, while most vary slightly in phase location, amplitude, and width to a lesser extent. Those small clusters are usually separated by 20--30 rotations, and this cluster pattern repeats roughly uniformly during the B or Bf-mode activity window. There is no clear discernible pattern concerning the MP's brightness, pulse width, or microstructure. The PC's microstructure behaviour seems mostly independent from the MP emission. The clusters of extensive PC microstructure and bright pulses in the MP (flaring) alternate, with the occasional alignment close in rotation space but without any obvious pattern. Similarly, there is a lack of correlation with the other single-pulse phenomena.

In summary, we find that the narrow emission components are indeed quasi-periodic shot-like micropulses with typical full-widths at half maximum between 0.2 and 0.4~ms and usual separations between 0.7 and 1~ms. The most prevalent quasi-periodicity occurs with a frequency near 1.2~kHz or a period of 0.83~ms. There are typically 20 micropulses in the combined PC and MP phase range.

\subsubsection{Mode mixing}
\label{sec:modemixing}

Secondly, the pulsar occasionally clearly shows low-level PC activity within long stretches of what appears to be otherwise pure Q-mode emission. We show examples of this in Fig.~\ref{fig:pulsemorphology} second row. In those pulses, only one or two very narrow (two to four phase bins) peaked emission components are visible above the baseline noise in the PC phase range 0.07--0.1. Their peak S/N is low compared with the MP emission, ranging from 0.1 to 0.7 of the maximum MP S/N. The pulses occur within the PC phase envelope and are sufficiently distant from the MP. Hence, they genuinely come from the PC and are unlikely due to bridge emission between the two profile components. In a subsequent step, we highlighted the rotations with low-level PC activity as identified from our visual inspection of the dynamic spectrum plots on the single-pulse stack plots (Fig.~\ref{fig:singlepulsestacks01}) with coloured markers (not shown in Fig.~\ref{fig:singlepulsestacks01} for readability). Guided by these markers, the PC activity becomes faintly visible in the pulse stacks. For instance, one can see low-level PC emission in the approximate pulse number ranges 274--318, 1232--1267, 1744--1893, 2508--2598, 3181--3258, and 3701--3724 in the first pointing on 2023-04-24 (Fig.~\ref{fig:singlepulsestacks01} left panel). In the second pointing (Fig.~\ref{fig:singlepulsestacks01} right panel), we see low-level PC activity in the rotation ranges 77--272, 1521--1612, and 1700--1715. That is, the PC emission stays active at a significantly reduced level for about 50--100 rotations after a mode switch from the B or Bf-mode to the Q-mode. On other occasions, the PC emission activates randomly during long Q-mode stretches several hundred rotations after a mode switch. This appears to happen preferentially near the middle of a given Q-mode stretch. The reverse is true as well. There are many rotations within otherwise pure B or Bf-mode emission stretches where the PC profile component is entirely absent.

Additionally, the pulse number range 3701--3724 in the first pointing shows another peculiar behaviour. The PC becomes active for several rotations in this region far from the previous mode switch. One might think that the pulsar briefly switches to the B or Bf-mode, but this is not the case. The MP lacks any noticeable change, the PC emission exhibits a lower intensity, smaller $N_\text{p}$, and is shifted to later pulse phases (0.1) compared with the typical B or Bf-mode PC contribution. The same is apparent from the feature timelines in Fig.~\ref{fig:featuretimeline}. Hence, this pulse number range might show an entirely new and distinct emission mode of the pulsar.

The low-level activity of the PC after a switch to the Q-mode and spuriously within long otherwise pure Q-mode stretches indicates clearly that the pulsar exhibits mode mixing. This is surprising, as it contradicts conventional pulsar knowledge \citep{2017Hermsen}. It seems that several modes can be simultaneously active in the pulsar, which also means that a binary classification (PC on vs.\ off) into disjoint modes is impossible. In other words, the currently observed emission state of the pulsar is likely a weighted sum of its (disjoint) emission modes, where the mixture weights determine the various contributions. For instance, during the Q-mode, the contribution that drives the PC emission might have an almost negligible weight, whereas during the B and Bf-mode, it might reach its maximum.

\subsubsection{Low-intensity square-like pulses}
\label{sec:weakpulses}

Thirdly, we discovered strangely looking, almost featureless pulses with very low S/N $\leq 5 - 10$ per phase bin. A less sensitive telescope might register them as nulling. Some of them have an almost square-like appearance (Fig.~\ref{fig:pulsemorphology} third row) with boxcar-equivalent widths between 0.01 and 0.03 phase (8--23~ms). The low-intensity pulses occur irrespective of mode; that is, they appear during both long Q-mode and B or Bf-mode pulse stretches. They often occur clustered in time with two to four low-intensity pulses after each other. They are pretty common in our single-pulse selection and the overall pulse stack, with relative occurrences between 30 and 40\%. We speculate that these low-intensity square pulses are produced by the most stable emission components in the pulsar's magnetosphere. They might resemble a stable baseline or plateau emission in the MP window. In principle, they might be evidence for a fourth emission mode of the pulsar, although our current Markov switching model implementation did not identify them as such.

\subsubsection{Flaring emission in the MP component}
\label{sec:flaring}

\begin{figure}
  \centering
  \includegraphics[width=\columnwidth]{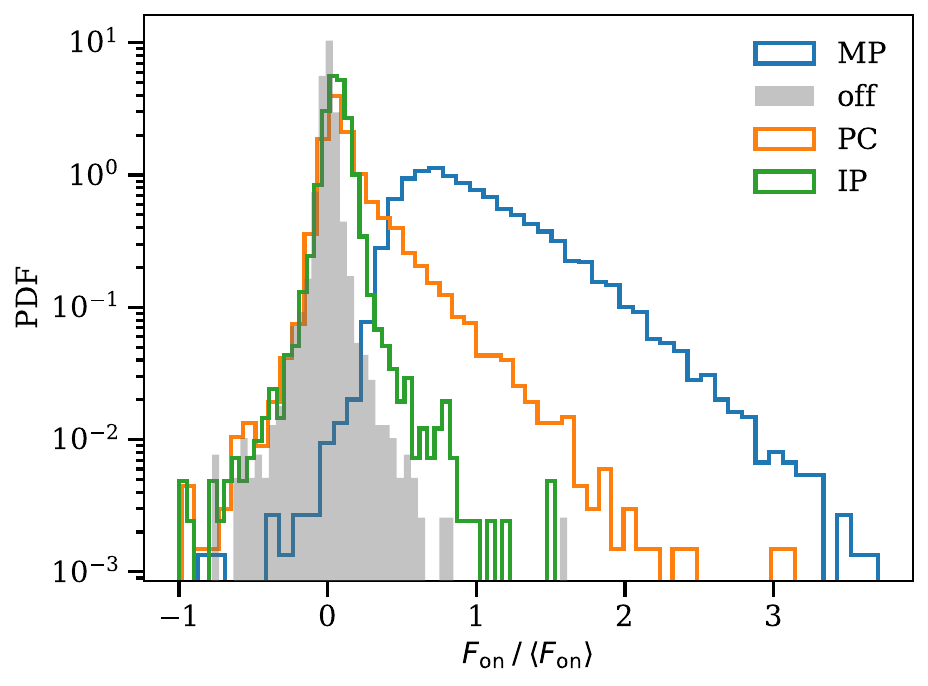}
  \caption{Mode-averaged pulse-energy distributions for each profile component (MP, PC, and IP) and the off-pulse baseline noise. The MP and PC distributions exhibit log-normal high-fluence tails and lack any obvious power-law behaviour.}
  \label{fig:penergydist}
\end{figure}

\begin{figure}
  \centering
  \includegraphics[width=\columnwidth]{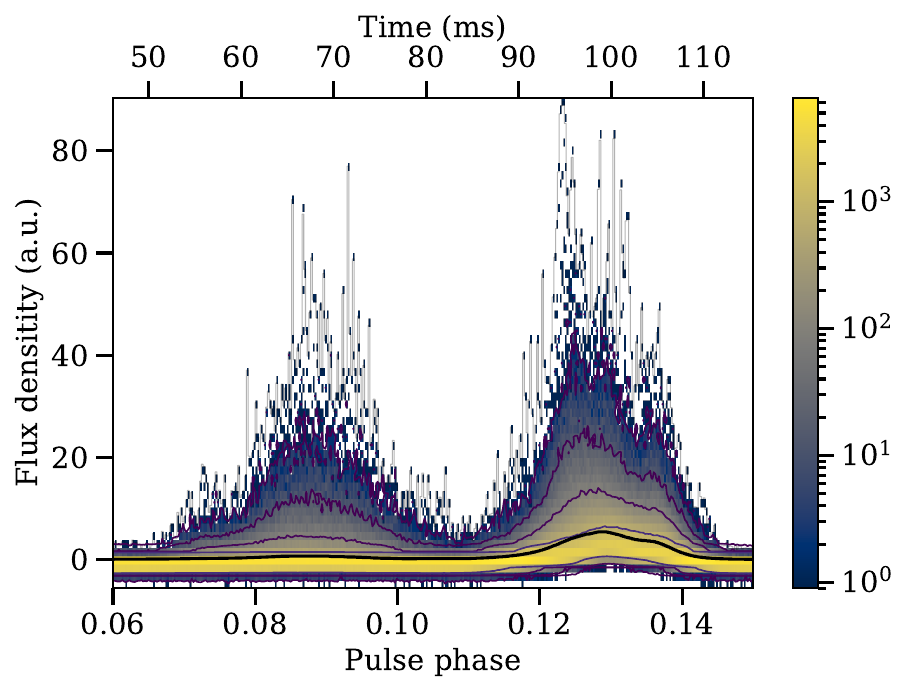}
  \caption{Phase-resolved single-pulse amplitude distribution of the PC and MP phase range. We show the mode-averaged 2D amplitude-phase distribution, that is, the number of pulses per flux density and phase bin, displayed on a logarithmic colour scale. The black solid line that peaks near 5~a.u.\ is the same average pulse profile as shown in Fig.~\ref{fig:pulseprofile}, the grey solid line represents the maximum amplitude, and the coloured solid lines are logarithmically spaced contours of the 2D distribution.}
  \label{fig:amphist}
\end{figure}

Moreover, the MP shows rare extremely bright bursts of emission of $\text{S/N} \geq 100 - 200$ per phase bin, always occurring in a specific pulse phase range near the centre of the MP window, during which the overall emission is dominated by that single peak (Fig.~\ref{fig:pulsemorphology} bottom row). There seems to be some interplay between the occurrence of the extremely bright pulses and the mode switching in the sense that the bright pulses typically happen shortly after a mode transition to the B or Bf-mode.

To investigate the bright pulses further, we measured the empirical pulse-energy (PE) distributions in separate phase ranges for each profile component (MP, PC, and IP) and the off-pulse baseline noise. We show the mode-averaged PE distributions referenced to a constant global mean on-pulse fluence $\left< F_\text{on} \right>$ in Fig.~\ref{fig:penergydist}. A log-normal distribution almost perfectly describes the MP PE distribution. The PC distribution is a mixture between a Gaussian at low and a log-normal distribution at higher fluences, likely representing the contributions from the Q-mode and the B or Bf-mode. The IP PE distribution differs only slightly from a normal distribution. There is a slight peak at $0.8 \left< F_\text{on} \right>$, and the distribution is positively shifted, appreciably above the off-pulse noise. The off-pulse distribution is well-described by a Gaussian, as expected from being primarily due to radiometer noise. The measured PE distributions lack any obvious power-law shape in a double logarithmic plot, besides a suggestive linear slope in the low-fluence PC distribution below $0.6 \left< F_\text{on} \right>$. Additionally, the high-fluence distribution tails show clear curvature in the same double logarithmic plot. Hence, the bright pulses are unlikely giant or micro-giant pulses. There is no evidence for this in our data set. The brightest pulses reach about $3.7$ (MP), $3.1$ (PC), and $1.5$ (IP) times the MP mean on-pulse fluence $\left< F_\text{on} \right>$ averaged over their respective phase ranges.

To understand where the bright pulses originate in the on-pulse envelope, we generated phase-resolved single-pulse flux density amplitude distribution plots for the entire data set, irrespective of mode. Figure~\ref{fig:amphist} shows the 2D amplitude-phase distribution in a logarithmic colour scale and contours. As is apparent, the brightest MP pulses occur predominantly at earlier phases near 0.123--0.125 at the leading edge of the MP. Additionally, there is a slightly lower peak near 0.129 phase, close to the MP's profile maximum. The brightest pulses exceed the average by a factor of 30 in intensity per phase bin. This agrees well with our visual inspection and the example plots shown in Fig.~\ref{fig:pulsemorphology}. The bright pulses seem to be created by flaring or bursting events near the leading edge of the MP, where the MP profile component widens slightly towards earlier phases. Those flaring events can even be seen in the pulse stacks in Figs.~\ref{fig:singlepulsestacks01}--\ref{fig:singlepulsestacks02}.

The situation is different for the PC, where the emission of narrow, bright micropulses occurs relatively uniformly across the PC phase window. There is a marginal clustering near 0.087--0.088 phase, near the PC's local profile maximum. The brightest PC pulses exceed its mode-averaged mean emission by a factor of 150.

\subsubsection{Pulse intensity fading after mode switches}
\label{sec:pulsefading}

Additionally, we noticed that the intensity of the pulses slowly increases or fades in after a mode switch to the B or Bf-mode and fades out after the switch back to the Q-mode. Besides the single-pulse profiles and dynamic spectra, one can also see this behaviour in the $F_\text{on}$ and S/N feature timelines in Fig.~\ref{fig:featuretimeline}, where the grey running mean lines help guide the eye. The pulsar seems to need some time to reach its maximum average radio intensity in that mode. The low-level PC activity in the Q-mode we discovered and described earlier likely contributes to the fading out of the pulsar's intensity after the switch.

\section{Discussion}
\label{sec:discussion}

\subsection{Mode switching behaviour and the Bf-mode}
\label{sec:bfmodediscussion}

PSR~B1822$-$09 has been known to exhibit two distinct emission modes (Q and B-mode) for a long time \citep{1981Fowler} and has been an example of a classical mode switching pulsar. Here, we presented strong evidence for the existence of at least a third distinct emission mode in the pulsar, which we called the bright flaring Bf-mode. It is characterised by a higher pulse-averaged flux density, more extreme sub-millisecond microstructure emission in the PC, and a significantly higher contribution of the PC to the pulsar's total flux density output. In particular, the fluence ratio between PC and MP is more similar than in the other modes, at around 1:2 (Fig.~\ref{fig:modecomparison}). Because of those properties, we consider it a flaring emission mode. In many regards, it is a more extreme version of the ordinary B-mode.

A closely related question is whether the Bf-mode simply represents the extreme end of the regular B-mode's properties. This does not seem to be the case. Our Markov switching model identified it confidently as a distinct emission mode. The separation from the B-mode is clear and most evident in the $N_\text{p}$ domain (Fig.~\ref{fig:markovmodelfitresults}). The data nicely separate into at least three distinct levels of $N_\text{p}$ centred on the coloured dashed lines with a spread given by the white-noise standard deviation $\sigma$. The fact that the $N_\text{p}$ time series consistently reaches the same three levels across our data set is the strongest evidence that these are indeed distinct modes.

Apart from the obvious differences in microstructure at the single-pulse level, the ratio of the component-integrated PC to MP fluence differentiates the three modes (Fig.~\ref{fig:modecomparison}). Interestingly, this is analogue to the amplitude ratio shape parameter $A_\text{pc} / A_\text{mp}$, where $A_\text{pc}$ is the amplitude of the PC and $A_\text{mp}$ that of the MP, which correlates with the pulsar's spin-down rate $\dot{\nu}$ on timescales of several years \citep{2010Lyne}. It might simply be that the pulsar, on average, switches more often into the flaring Bf-mode during its times of high-$\dot{\nu}$ (Fig.~5 of \citealt{2010Lyne}). Conversely, more frequent switching to the Bf-mode on average results in higher energy output, which in turn causes the star to spin down faster, that is, a higher dipole magnetic braking and higher $\dot{\nu}$. In our Markov switching model for the pulsar's moding, this behaviour can naturally be explained by a slight increase in the transition probability $p_{21}$ from the B-mode to the Bf-mode or potentially also directly from the Q-mode (Fig.~\ref{fig:markovstatediagram}). This would result, on average, in a higher presence of the Bf-mode. However, what physical mechanism could alter the pulsar's intrinsic state transition probabilities is not obvious. For instance, one could invoke an external perturbation of the pulsar's magnetosphere by an asteroid or other dense material from an accretion disc. In this picture, the transition probability might decay over several years to its original value as the system slowly returns to its original (unperturbed) state. This hypothesis is consistent with the observed return of the pulsar to its original $\dot{\nu}$ value.

Regarding pulsar energetics, an increase in radio energy output alone is insufficient. The excursions in $\dot{\nu}$ have roughly $\Delta \dot{\nu} \simeq 3 \%$ (Fig.~5 of \citealt{2010Lyne}), and thus the spin-down luminosity must increase by the same amount, $\Delta \dot{E} \simeq 3 \%$. A pulsar with $\dot{E} \simeq 4.5 \times 10^{33} \: \text{erg} \: \text{s}^{-1}$ might have a radio efficiency $\xi_\text{r} = 10^{-5} - 10^{-3}$ \citep{2014Szary}. We know from observations that the X-ray emission does not participate significantly in the mode switching in this pulsar \citep{2017Hermsen}. A $\gamma$-ray pulsar with comparable $\dot{E}$ to PSR~B1822$-$09 has a $\gamma$-ray efficiency around $\xi_\gamma = 5 \%$ with a large uncertainty \citep{2023Smith}. So, in principle, an increase in $\gamma$-ray energy output could account for the $3 \% \: \Delta \dot{E}$. However, B1822$-$09 is only faintly detected in GeV $\gamma$-rays, which suggests that its $\xi_\gamma$ is significantly lower than $5 \%$ or that we only see a small part of its $\gamma$-ray emission. To resolve the problem, one could invoke an unseen wind of energetic particles that carries away the spin-down luminosity difference and might switch in tandem with the radio emission.

\subsection{Mode mixing}
\label{sec:modemixingdiscussion}

We found evidence for mode mixing, that is, the presence of clear PC activity within long otherwise pure Q-mode stretches (Fig.~\ref{fig:pulsemorphology}) and the complete absence of PC emission in some otherwise pure B or Bf-mode stretches. This challenges the traditional assumption that the pulsar emission modes are entirely disjoint and presents complications for modelling the mode switching process, including our Markov switching model. Our findings agree well with those of \citet{2012Latham}, who called the phenomenon `parallel moding'. In addition to the pure Q and B-mode emission, they saw pulse sequences with B-mode behaviour with IP emission (BwIP), Q-mode behaviour with PC illumination (QwPC), and a third unspecified mode mixing type at the Q and B-mode transitions. As described above, we also see the QwPC behaviour. However, instead of taking the presence of IP emission as the characterising feature for the BwIP case, we used the absence of the PC, as it is a more obvious and reliable indicator given the IP's low flux density and limited change between the modes (Fig.~\ref{fig:modecomparison}). Other than that, we also see the mode mixing within long B or Bf-mode stretches.

\subsection{Fluctuation properties}
\label{sec:fluctuationpropertiesdiscussion}

Our results for the $P_3 = (41.7 \pm 0.9) \: P_1$ Q-mode modulation feature are in good agreement with the literature. In particular, \citet{2019Yan} reported a periodicity of $P_3 = (42.6 \pm 0.2) \: P_1$ at 1.4~GHz. \citet{2010Backus} found $P_3 = (43.75 \pm 1.0) \: P_1$ at 325~MHz. Earlier work suggested a $P_3 \approx 40 \: P_1$ at 1.4~GHz \citep{1994Gil} and 1.6~GHz \citep{1981Fowler}. Those values are entirely consistent within the uncertainties. However, there is a slight $\sim$3~$\sigma$ tension with the measurement of $P_3 = (46.55 \pm 0.88) \: P_1$ at 325~MHz reported by \citet{2012Latham}. We believe that most of the variation can be attributed to how the maximum Fourier power bin was determined in the LRFS and the presence of spectral leakage if not accounted for by windowing the FFT. Overall, the $P_3$ Q-mode modulation is stable in time over more than 40 years and a factor of five in observed radio frequency.

Interestingly, the longer period fluctuation features we found at $52.6$ and $66.7\: P_1$ have not been reported in the literature. They are of lower Fourier S/N, but clearly above $5~\sigma$ and distinct from the $41.7 \: P_1$ feature in the LRFS (Fig.~\ref{fig:lrfs}). We suspect that our long Q-mode data stretch (1000 rotations), absence of RFI, and excellent data quality enabled their detection.

Our analysis indicates that the Q-mode fluctuation is primarily or solely due to longitude-stationary amplitude modulation. There is no evidence of sub-pulse drifting. This agrees with previous work at 1.4~GHz \citep{1994Gil, 2019Yan}, but is in tension with the analyses performed at 325~MHz that suggested a phase or amplitude-phase modulation \citep{2010Backus, 2012Latham}. This might suggest a transition in Q-mode fluctuation character from pure phase or mixed amplitude-phase modulation at low frequencies to primarily amplitude modulation at 650~MHz and above. Alternatively, rapid diffractive interstellar scintillation of the pulsar's radio signal could potentially have induced extrinsic phase modulation in earlier work if the pulsar's radio spectral index varied considerably across the MP envelope and the instrumental band was small compared with the scintillation bandwidth, that is, due to insufficient spectral averaging of older narrow-band data.

We found no evidence of modulation in any of the profile components in our B and Bf-mode data. In particular, there was no noticeable modulation at the previously reported periods of $P_3 = 11 \: P_1$ \citep{1981Fowler, 1994Gil} or $P_3 = 70 \: P_1$ \citep{2012Latham}.

\subsection{Difference between Q-mode amplitude modulation and mode switching}

\citet{2019Yan} discussed the Q-mode modulation in the context of longitude-stationary amplitude modulation and periodic nulling. They concluded that it is the former that happens. Indeed, we did not find any evidence for nulling in our data set. The pulsar emitted continuously during each rotation that was unaffected by RFI, as we show in Figs.~\ref{fig:singlepulsestacks01}--\ref{fig:singlepulsestacks02}, \ref{fig:qmodeampmodulation}, and \ref{fig:componentcorrelation}. Hence, it is not nulling that is responsible for the modulation. Another possibility is that subtle mode switches cause the modulation. Our hidden Markov model considered all low $N_\text{p}$ emission to originate from the same mode, that is, the Q-mode (Fig.~\ref{fig:markovmodelfitresults}). It refrained from decomposing the Q-mode modulation into further disjoint modes. In fact, a hint of Q-mode modulation is visible in the $N_\text{p}$ time series (Figs.~\ref{fig:featuretimeline}--\ref{fig:markovmodelfitresults}). However, it is fully explained as autoregressive behaviour with positive $\phi_1$ and noise. Given our current modelling technique and available data, we conclude that the Q-mode modulation is longitude-stationary amplitude modulation and a separate phenomenon from mode switching.

\subsection{Q-mode internal phase-locking versus asynchronous mode switching}
\label{sec:qmodephaselockingdiscussion}

Our data nicely demonstrate the phase-locked mode switching between the pulsar's MP, PC, and IP profile components. The PC and MP switch fluence synchronously to each other (positively correlated), while both switch asynchronously (anti-correlated) to the IP (Fig.~\ref{fig:componentcorrelation}). This is a well-known but poorly understood phenomenon \citep{1982Fowler, 2005DyksA}. Additionally, we found that the MP and IP emission are phase-locked at a lower level during the pulsar's Q-mode (Fig.~\ref{fig:qmodeampmodulation}). Both the MP and IP exhibit the same amplitude modulation periodicity, $P_3 = (41.7 \pm 0.9) \: P_1$, and are positively correlated, meaning synchronous fluence modulation. Interestingly, the correlation behaviour within the Q-mode seems to be reversed compared with the overall mode switching.

The phase offset of the low-level phase-locking between the MP and IP during the Q-mode is important and has been debated. For instance, \citet{2010Backus} used longitude-longitude correlation maps and found a significant (asymmetric) correlation of the IP delayed by $2 P_1$ with the leading region of the MP. \citet{2012Latham} reported a phase offset of about $12 P_1$ in modulation behaviour using modulation period $P_3$-folds and longitude-longitude correlation maps. \citet{2019Yan} measured the cross-correlation between the pulse energy time series of the IP and the leading component of the MP. They found the strongest correlation at zero lag, that is, between the IP and the directly following MP. Our analysis agrees with \citet{2019Yan} that there is zero phase offset between the fluence modulation patterns of the leading MP profile sub-component and the IP, meaning that they are perfectly in phase. There is no obvious feature in the MP-IP cross-correlation at a lag of $2 P_1$. The cross-correlation falls off smoothly from its maximum at zero lag and reaches almost precisely zero at a lag of $12 P_1$. This might be purely coincidental, but it could relate to what \citet{2012Latham} saw in their $P_3$-folds.

\subsection{Peculiar single-pulse phenomena}

At least five peculiar single-pulse phenomena are happening in PSR~B1822$-$09: (1) extensive quasi-periodic microstructure in the PC, (2) PC activity within the Q-mode (mode mixing or instantaneous switches), (3) square-like almost null pulses, (4) extremely bright pulses, and (5) pulse intensity fading in after mode switches. Only some of these have been described in the literature.

\subsection{Quasi-periodic profile microstructure and its frequency dependence}
\label{sec:microstructurediscussion}

\citet{1994Gil} studied the pulsar's profile microstructure near 1.4~GHz and reported the presence of many narrow quasi-periodic microstructure spikes of about 150 to 300~$\mu \text{s}$ width in its PC. The MP had a more amorphous structure with typical microstructure widths of 600~$\mu \text{s}$. This agrees well with our measurements at 650~MHz. Their single-pulse profile plots look qualitatively very similar to ours (Figs.~\ref{fig:pulsemorphology}--\ref{fig:pcmicrostructure}), and the full-widths at half-maximum match closely (200--400~$\mu \text{s}$). The typical separations were between 0.7 and 1~ms with a most-prevalent quasi-periodicity of 0.83~ms in our data. The microstructure of the PC and MP also differ drastically at 650~MHz (Fig.~\ref{fig:pcmicrostructure}). The reported 150~$\mu \text{s}$ minimum width strengthens our conclusion that we resolved the majority of the micropulses in our analysis. However, the native time resolution of our data will allow for a more detailed investigation in the future. Interestingly, the width of the micropulses seems to be independent of radio frequency, at least at this time resolution. This would be curious if the PC followed the usual radius-to-frequency mapping model with the pulse width broadening towards lower frequencies. However, it strangely does not \citep{1994Gil}. If the observed microstructure is representative of sparks in the pulsar's magnetosphere, their typical projected size must have stayed constant over roughly 30 years and varied little with radio frequency.

\subsection{Pulsar emission geometry}
\label{sec:emissiongeometrydiscussion}

While there is no clear answer to the pulsar's overall emission geometry, particularly the emission site of its PC profile component remains a mystery \citep{2012Latham}. The PC strangely defies the standard radius-to-frequency mapping model, and its pulse width stays more or less constant between about 100~MHz and 11~GHz. This is in contrast to the IP and MP, which exhibit a strong and mild frequency dependence, respectively. The PC-MP separation stays approximately constant over the same frequency range \citep{1994Gil}. Given the above, we speculate that the PC is emitted at an altitude different from both the MP and IP. The magnetic field line curvature in the radial direction along the pulsar emission cone must be low at its emission site so that there is almost no radius-to-frequency mapping and, thus, virtually no observable frequency dependence of its pulse width. For instance, it might arise from emission from higher up in the pulsar's magnetosphere, which gets shifted to earlier pulse phases by aberration and retardation effects. The resulting observed phase shift $\Delta \phi$ relates to the difference in emission height $\Delta h_\text{em}$ as
\begin{equation}
    \Delta h_\text{em} = \Delta \phi \: \frac{ c P_1 }{ 4 \pi },
\end{equation}
where $c$ is the speed of light \citep{2007Weltevrede}. We measured a PC-MP separation of about 0.041~phase or 14.8~deg peak-to-peak in our 650~MHz data (Fig.~\ref{fig:pulseprofile}). Thus, if the PC's location is determined by aberration and retardation effects, its emission height must be offset by roughly 753~km from that of the MP, which corresponds to $\sim$2\% of the light cylinder radius $R_\text{LC}$. This is reasonable and fits within the 1--5\% $R_\text{LC}$ emission height range typical for radio pulsars \citep{2017Hermsen}. It also agrees with recent rotating vector modelling work near 1.3~GHz \citep{2023Johnston}, as long as the MP (and IP) emission height is less than $\sim$250~km so that the PC emission height is $\leq 1000~\text{km}$. Hence, we believe that the PC originates from slightly higher altitudes above the same polar cap that produces the MP emission and that the PC and MP on the one hand and the IP on the other come from opposite magnetic poles. The former is in line with previous work \citep{2007Weltevrede, 2012Latham} and the latter is the most widely accepted belief \citep{1986Hankins, 1994Gil, 2010Backus, 2017Hermsen}. Alternatively, the PC and MP emission might come from the same emission altitude but from particles with significantly different Lorentz factors.

\subsection{Magnetar-like precursor component characteristics}

PSR~B1822$-$09's PC component shows several emission characteristics reminiscent of radio-loud magnetars. It has an unusually flat spectral index \citep{1981Fowler, 1994Gil}, a linear polarisation fraction close to 100\% \citep{1980Manchester, 2018Johnston}, strong variability, and exhibits spiky emission with extremely narrow (0.2--0.4~ms) and intense quasi-periodic shot-like bursts, which often exceed the MP flux density per bin. In contrast, the MP behaves more like an ordinary radio pulsar pulse, except for the longitude-stationary amplitude modulation around the MP's leading profile hump. Thus, it seems that the pulsar combines both magnetar and canonical radio pulsar attributes, which is rarely seen in the known pulsar population.

\subsection{Distinguishing pulsar-intrinsic and propagation effects}

Our analyses that are based on the pulsar's flux density, such as the fluctuation spectra and the pulse-energy distributions, naturally measure a combination of pulsar-intrinsic properties that directly probe the pulsar radio emission mechanism(s) and effects due to the propagation of the pulsar's radio signal through the turbulent ionised interstellar medium. Analyses, such as our mode classification and separation, that take into account other single-pulse profile features (e.g.\ the number of peaked components $N_\text{p}$) are much less affected by scintillation above a limiting S/N or flux density threshold. We estimated the effect of scintillation in the following. We computed a scatter broadening time $\tau_s = 0.27~\mu \text{s}$ at 650~MHz towards the pulsar using the \texttt{YMW16} Galactic free-electron model \citep{2017Yao}. This equates to a diffractive scintillation bandwidth $\Delta \nu_s \approx 1.16 / (2 \pi \: \tau_s) = 0.68~\text{MHz}$. This is well above the native frequency resolution of our data ($\sim$0.1~MHz), meaning that we could indeed resolve scintles and study the pulsar's scintillation in a future analysis. We then calculated the expected modulation indices due to strong diffractive (DISS) and refractive interstellar scintillation (RISS) and the total modulation index following \citet{2018Jankowski}. There is practically zero time averaging of the single-pulse data, only across the observed effective bandwidth. The resulting modulation indices are $m_\text{DISS} = 0.16$, $m_\text{RISS} = 0.32$, and $m_\text{tot} = 0.42$. The flux density modulation is dominated by the slowly time-varying refractive scintillation part, which is mostly irrelevant for this work. All our measured modulation indices exceed $m_\text{tot}$ (Figs.~\ref{fig:singlepulsestacks01}--\ref{fig:singlepulsestacks02}, and \ref{fig:qmodeampmodulation}--\ref{fig:lrfs}), often drastically. This means that propagation effects are negligible for our conclusions in this paper, particularly the mode classification and that pulsar-intrinsic phenomena dominate any flux density modulation.

\section{Conclusions}
\label{sec:conclusions}

PSR~B1822$-$09 is an interesting pulsar that exhibits complex behaviour at the single-pulse level. We have presented its single-pulse stacks and integrated profile at 650~MHz, investigated its mode changing properties using a novel hidden Markov switching model, applied a fluctuation spectra analysis to the data, and discussed its single-pulse profile morphology. Regarding the profile morphology in particular, we focused on five peculiar phenomena found in our data. We have significantly added to the already extensive literature about this pulsar and confirmed several of its properties while refuting others. While doing so, we introduced several new analysis techniques to the pulsar domain. This work demonstrates what is possible with a small subset of the SUSPECT project data.

\section{Data availability}
\label{sec:dataavailability}

The high-level data underlying this article will be shared on reasonable request to the corresponding author.

\begin{acknowledgements}

We thank Killian Lebreton for his contributions to this work as part of his Master's degree project at the Universit\'{e} d'Orl\'{e}ans, the Nan\c{c}ay--Orl\'{e}ans pulsar group for helpful discussions, Yogesh Maan for clarifications about uGMRT data reduction, the reviewer for constructive questions and comments, and the staff of the GMRT who have made these observations possible. We acknowledge the use of the Nan\c{c}ay Data Centre computing facility (CDN -- Centre de Donn\'{e}es de Nan\c{c}ay). The CDN is hosted by the Observatoire Radioastronomique de Nan\c{c}ay in partnership with Observatoire de Paris, Universit\'{e} d'Orl\'{e}ans, OSUC and the CNRS. The CDN is supported by the R\'{e}gion Centre Val de Loire, d\'{e}partement du Cher. The GMRT is run by the National Centre for Radio Astrophysics of the Tata Institute of Fundamental Research. This work has been supported by ANR-20-CE31-0010.

\end{acknowledgements}


\bibliographystyle{aa} 
\bibliography{references} 



\listofobjects

\end{document}